\documentclass[aip,jcp,preprint]{revtex4-1}
\usepackage{graphicx}% Include figure files
\usepackage{dcolumn}% Align table columns on decimal point
\usepackage{bm}% bold math
\usepackage{amsmath}
\usepackage{amsfonts}
\usepackage{amssymb}
\usepackage{braket}
\usepackage{subcaption}
\usepackage{xcolor}
\usepackage{soul}

\renewcommand{\hl}[1]{#1} % disactivate highlight

\let\oldsqrt\sqrt
\def\sqrt{\mathpalette\DHLhksqrt}
\def\DHLhksqrt#1#2{%
\setbox0=\hbox{$#1\oldsqrt{#2\,}$}\dimen0=\ht0
\advance\dimen0-0.2\ht0
\setbox2=\hbox{\vrule height\ht0 depth -\dimen0}%
{\box0\lower0.4pt\box2}}

\begin{document}

\preprint{A16.03.0268}

\title{Simulation of Entangled Polymer Solutions}% Force line breaks with \\
%\thanks{Footnote to title of article.}

\author{Airidas Korolkovas}
\email{korolkovas@ill.fr}
\affiliation{Institut Laue-Langevin, 71 rue des Martyrs, 38000 Grenoble, France}
\affiliation{Universit\'{e} Grenoble Alpes, Liphy, 140 Rue de la Physique, 38402 Saint-Martin-d'H\`{e}res, France}
\author{Philipp Gutfreund}
\affiliation{Institut Laue-Langevin, 71 rue des Martyrs, 38000 Grenoble, France}
\author{Jean-Louis Barrat}
\affiliation{Institut Laue-Langevin, 71 rue des Martyrs, 38000 Grenoble, France}
\affiliation{Universit\'{e} Grenoble Alpes, Liphy, 140 Rue de la Physique, 38402 Saint-Martin-d'H\`{e}res, France}

\date{\today}% It is always \today, today,

\begin{abstract}
We present a computer simulation of entangled polymer solutions at equilibrium. The chains repel each other via a soft Gaussian potential, appropriate for semi-dilute solutions at the scale of a correlation blob. The key innovation to suppress chain crossings is to use a pseudo-continuous model of a backbone which effectively leaves no gaps between consecutive points on the chain, unlike the usual bead-and-spring model. Our algorithm is sufficiently fast to observe the entangled regime using a standard desktop computer. The simulated structural and mechanical correlations are in fair agreement with the expected predictions for a semi-dilute solution of entangled chains.
\end{abstract}

%Use showkeys class option if keyword
\maketitle

\section{Introduction}
Simulation of entangled polymer solutions is a long standing challenge in the field of macromolecules. While it is possible to simulate polymer melts of sufficiently long chains where entanglement effects become visible, achieving a comparable result in polymer solutions remains evasive, despite ongoing increase in computer speed and algorithm efficiency. To start with, entanglement is not an interaction \emph{per se} that one could insert in the simulation code. Rather, it is an emergent phenomenon due to the uncrossability of very long, interpenetrating polymer chains. The main challenge from the numerical point of view is to resolve chain motion with sufficient accuracy for there to be no crossings over a time span exceeding the one required for a chain to diffuse a distance equal to its own size.

Most of the previous simulation effort on entanglement was geared for polymer melts rather than their solutions. A popular model by \citeauthor{kremer1990}\cite{kremer1990} (KG) designed for melts is based on hard, impenetrable beads tightly bound by stiff nonlinear springs. The beads are often modeled by the steep repulsive part of the Lennard-Jones potential, also known as the Weeks-Chandler-Andersen (WCA) potential, whereas the connectivity is enforced by finitely extensible nonlinear elastic (FENE) springs. Alternatively, an even better barrier against crossings is obtained in lattice-based simulations\cite{dorgan2012}, with the downside that the chain conformation is unrealistically limited to only a handful of coordinations which depend on the arbitrary choice of the lattice (cubic, face-centered cubic, etc.)

\begin{figure}[bht]
  \begin{subfigure}[b]{0.4\textwidth}
    \includegraphics[width=\textwidth]{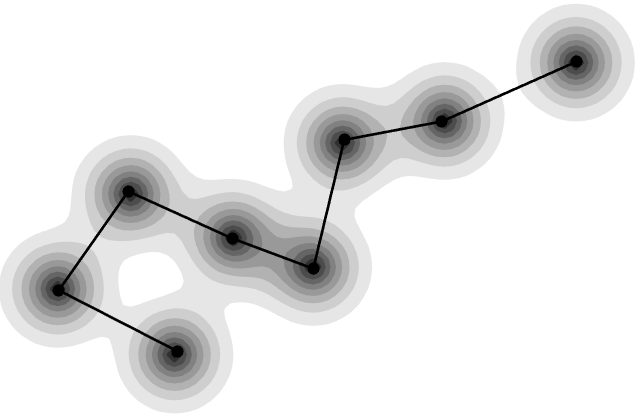}
    \caption{Bead-and-spring model: $J=N=8$.}
    \label{fig:1}
  \end{subfigure}
  \begin{subfigure}[b]{0.4\textwidth}
    \includegraphics[width=\textwidth]{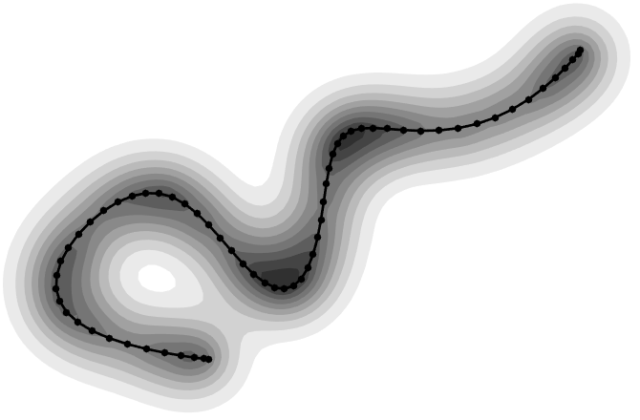}
    \caption{Pseudo-continuous model: $J=8N=64$.}
    \label{fig:2}
  \end{subfigure}
	\caption{Two models of the same molecule with $N$ degrees of freedom but different number of particles $J$. Shading denotes the repulsive potential $\Phi$ of range $\lambda$.}\label{model}
\end{figure}

In a polymer melt the entanglement length is typically within a factor of 10-100 longer than the natural interatomic distances, which is not very far apart and so the KG method is an adequate approach to simulate the liquid. However, if we dilute the system by adding solvent, while at the same time increase molecular weight to maintain a high degree of chain interpenetration (the semi-dilute regime), the computational time becomes a major issue. The entanglement length is now much greater than the interatomic scale, and the rate of chain collision is many orders of magnitude lower than the timescale required to follow the hardcore WCA+FENE interactions. Often in applications we want to focus on the physics of entanglement and we are less interested in the small features on the atomic scale. The main strategy is hence to simulate the polymers at a coarser scale\cite{padding2011systematic,masubuchi2014}, which basically means softer beads and looser springs. Alas, this quickly opens up gaps along the backbone (see Figure~\ref{fig:1}), through which the chains can cross each other and the entanglement behavior is lost.

To mimic the effects of entanglement, several recent studies have introduced temporary \emph{attractive forces}, called slip-springs\cite{chappa2012, langeloth2013, uneyama2012, hernandez2015} or slip-links\cite{masubuchi2001, hernandez2013}, between nearby beads. As an extreme example\cite{kindt2007}, one can replace the entire chain by just a single particle at the expense of having to invent and justify effective entanglement interactions with other such ``particles''. 

A rival camp of thought introduces additional \emph{repulsive forces}. One suggestion is to topologically detect the segments which have crossed during the time step, and then repel them back using the Twentanglement\cite{padding2001} algorithm. On second thought, why bother with topology at all instead of simply repelling the nearby segments even before they had a chance to cross, using the so-called segmental repulsive potential (SRP)\cite{pan2003, lahmar2009, yamanoi2011}?

In our recent work\cite{korolkovas2016simulating} we briefly mentioned a model which takes the SRP strategy even further and completely blurs the distinction between ``bead'' and ``segment''. The present paper explores this idea in much greater detail. The chain in theory is a fully continuous curve with $N$ degrees of freedom, which is discretized for computational purposes by drawing as many samples $J\gg N$ as needed such that the distance between consecutive points $|\mathbf{R}_j-\mathbf{R}_{j-1}|\ll \lambda$ is much smaller than the range of the excluded volume force, as shown in Figure~\ref{fig:2}.  A soft Gaussian potential is perfectly adequate to repel such pseudo-continuous chains, whereas a linear Hookean spring interaction keeps them connected. The time evolution is described by a stochastic first order equation of motion known as the Brownian thermostat. The random force is truncated at high frequencies, which reduces its peak amplitude, thus making chain crossings even less likely.

One alleged disadvantage is that we end up with a very dense and computationally demanding $N$-body (or rather, $(J\gg N)$-body) problem. To mitigate this issue, in Section~\ref{force} we propose an approximate algorithm which uses two staggered grids and splits the Gaussian potential into its short- and long-range contributions, each of which is very fast to evaluate. The code is highly parallel and is straightforward to further accelerate using GPU computing.

To validate our algorithm we have performed a series of computer simulations in the regime which can be mapped to semi-dilute polymer solutions. The obtained chain trajectories were analyzed to determine various structural and mechanical correlations. In particular, the self-diffusion coefficient scaled as $D\propto N^{-2}$, and the longest relaxation time scaled as $\tau_d \propto N^3$ for long $N>256$ chains. While our model does not explicitly prevent chain crossings, it does suppress them sufficiently for the reptation behaviour to emerge, thus reaching a fair agreement with well-known experimental and theoretical facts.

\section{The continuous model}
\begin{figure}[bht]
\begin{minipage}{.49\textwidth}
	    \caption{Typical state of semi-dilute polymer in two dimensions. The chains have $N=16$ degrees of freedom and keep a distance of about $\lambda$ from each other.}\label{blobs}
 \includegraphics[width=0.88\linewidth]{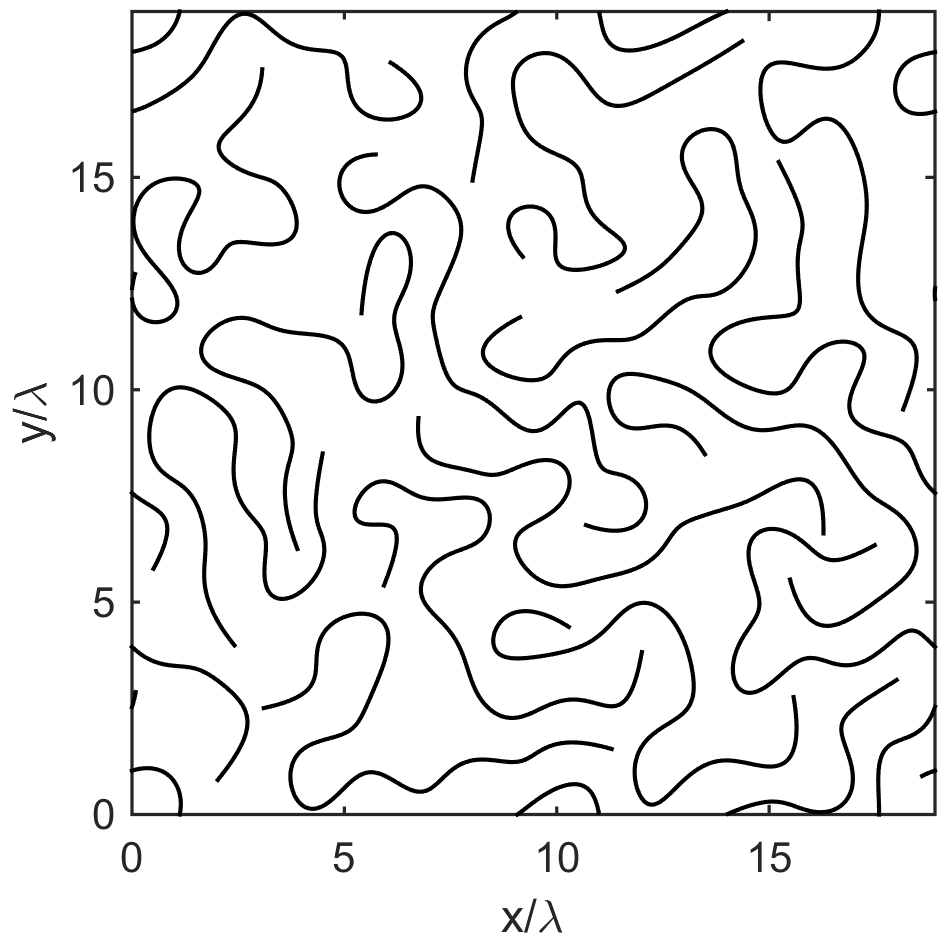}
\end{minipage}%
\hfill
\begin{minipage}{.49\textwidth}
	    \caption{$C=16$ chains with $N=32$ degrees of freedom in three dimensions. Yellow color indicates the tension $|\partial \mathbf{R}/\partial s|^2$, see Eq.~\eqref{Hedwards}.}\label{3D}
			\begingroup
			\sbox0{\includegraphics{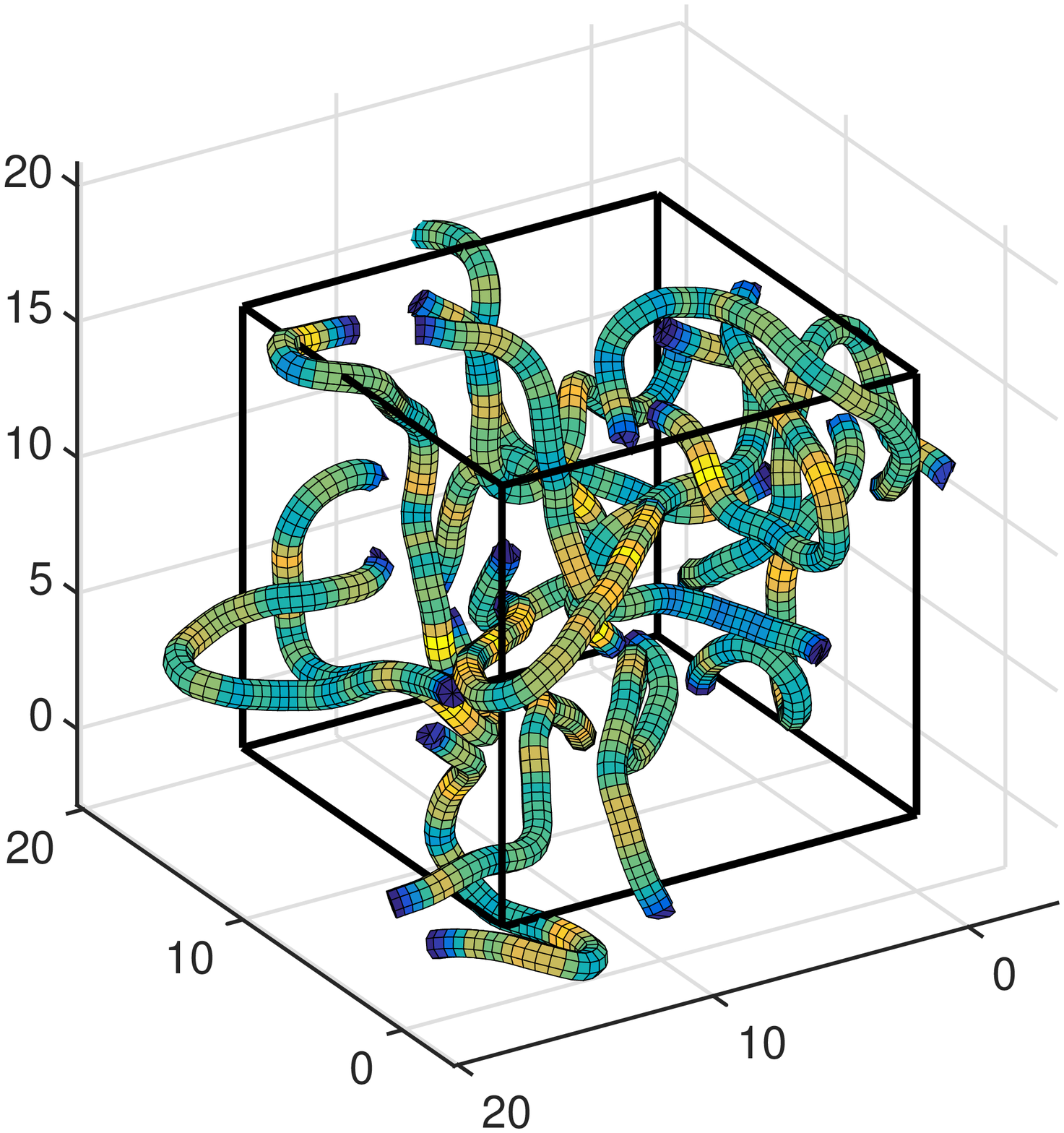}} % measure the original image size
			\includegraphics[clip,trim={.25\wd0} {.05\ht0} {.25\wd0} {.1\ht0},width=\linewidth]{./fig/3Dview.eps} % left, bottom, right, top
		\endgroup
\end{minipage}
\end{figure}
We consider a semi-dilute polymer solution consisting of a number $C$ of chains in a volume $V$, with periodic boundary conditions as shown for clarity in Figure~\ref{blobs} for the two dimensional case. All subsequent calculations will be carried out in the three dimensional case, illustrated in Figure~\ref{3D}. Each chain is nominally composed of $N$ monomers, or blobs, linked by a linear spring interaction of length $b$. We start with the continuous coil, or Edwards model\cite{edwards}, where the $c^{\text{th}}$ chain at time instant $t$ is represented by a continuous path $\mathbf{R}_c(t,s)$ with the monomer label $s\in(0,1)$. The energy of a configuration is
\begin{equation}\label{Hedwards}
H = \frac{3k_BT}{2N b^2} \sum_{c=1}^C \int_0^1  ds\, \left|\frac{\partial \mathbf{R}_c}{\partial s}\right|^2  + \frac{N}{2}\sum_{c=1}^C\sum_{c'=1}^C \int_0^1 \int_0^1ds\, ds'\, \Phi[\mathbf{R}_c(s) -\mathbf{R}_{c'}(s')].
\end{equation}
The first term is the attractive spring interaction of entropic origin,\cite{de1979scaling} while the function $\Phi$ denotes the excluded volume interaction, with $v\approx \lambda^3$ being the excluded volume parameter:
\begin{equation}\label{gaussian}
\Phi(\mathbf{r}) = vN k_B T \delta (\mathbf{r}) \approx N k_B T \exp \left(-\frac{\mathbf{r}^2}{2\lambda^2}\right).
\end{equation}
The Dirac delta $\delta(\mathbf{r})$ approximation is commonly used in continuum theories, while the Gaussian approximation with its finite range $\lambda \approx b$ is more suited for numerical calculations\cite{bolhuis2001}.

At equilibrium, the model can be most readily identified with a semi-dilute polymer solution at density $\rho$ and molecular weight $M_w$. Scaling theory\cite{de1979scaling} predicts the number of blobs and their size to scale as
\begin{equation}\label{mapping}
N \propto \rho^{5/4}M_w \quad \text{and} \quad \lambda \propto \xi \propto \rho^{-3/4}.
\end{equation}
This mapping is valid for semi-dilute solutions $\rho^* \ll \rho \ll \rho^{**}$, but could also be extended to melts, provided that the correlation length $\lambda(\rho_{\text{melt}}) \gg b_0$ is substantially greater than the size of an atom $b_0$, which may be a reasonable assumption for some chemical species. A blob particle contains both the polymer and the associated solvent, so we do not add explicit solvent particles.

Far from equilibrium, such as under a strong shear flow, or just in general whenever the chains are highly stretched as in a polymer brush, the above mapping breaks down. The simulation can still be performed, but one will be obliged to use more blobs $N>N_{\text{eq}}$ and a sharper potential $\lambda < \lambda_{\text{eq}}$ until eventually the atomic scale is reached and one may as well switch back to a Kremer-Grest type of approach.

In a semi-dilute solution the hydrodynamic interactions are screened and are not important\cite{deGennes1976} for distances beyond $\xi \approx \lambda$, and therefore are not included in the model. The chain dynamics can then be described by the stochastic Brownian equation of motion:
\begin{equation}\label{brownian}
\zeta \frac{\partial \mathbf{R}_c(t,s)}{\partial t} = \left(\frac{3k_B T}{Nb^2}\right) \frac{\partial^2 \mathbf{R}_c(t,s)}{\partial s^2}  -N\nabla U(\mathbf{r})_{\mathbf{r}=\mathbf{R}_c(t,s)} + \sqrt{2k_B T \zeta}\mathbf{W}_c(t,s)
\end{equation}
where $\zeta = 6\pi \eta_s b N$ is the friction coefficient of the center of mass, $\mathbf{W}_c(t,s)$ is the Wiener process satisfying $\braket{\mathbf{W}_c^{\alpha}(t,s) \mathbf{W}_{c'}^{\beta}(t',s')} = \delta^{\alpha \beta} \delta_{cc'} \delta(t-t')\, \delta(s-s')$, and
\begin{align}
U(\mathbf{r}) &= \sum_{c=1}^C \int_0^1 ds\, \Phi[\mathbf{r}-\mathbf{R}_c(s)] \\
& \approx \sum_{c=1}^C \sum_{j=1}^{J} \Phi_0 (\mathbf{r} - \mathbf{R}_{c,j})
\end{align}
is the total excluded volume field. The natural time unit is the microscopic Rouse time
\begin{equation}
\tau = \frac{6\pi \eta_s b^3}{k_B T},
\end{equation}
which is roughly the time it takes one blob to diffuse a distance equal to its own size. In contrast, the momentum relaxation time
\begin{equation}
\tau_m = \left(\frac{m}{6\pi \eta_s b} \approx \frac{\rho_0 b^2}{\eta_s}\right) \ll \left(\tau \approx \frac{\eta_s b^3}{k_B T}\right)
\end{equation}
would be the time during which the thermal velocity $\braket{\mathbf{v}^2} = 3k_B T/(2m)$ of the coarse particle $\lambda \approx b$ decorrelates significantly from its initial value. The particle mass $m$ is assumed to contain both the polymer and the surrounding solvent molecules, so the density $\rho_0 \approx 1\, \text{g/cm}^3$ refers to the overall density of the liquid. The Brownian equation~\eqref{brownian} of motion is justified as long as the above inequality $\tau_m \ll \tau$ holds and we can ignore inertia. In terms of physical polymer density $\rho$ and using Equation~\eqref{mapping}, the inequality can be expressed as
\begin{equation}
\rho \ll \rho_0 \left(\frac{b_0 \eta_s^2}{\rho_0 k_B T}\right)^{4/3} \approx \rho^{**}
\end{equation}
where $b_0 \approx 1\, \text{nm}$ is the size of the physical monomer. Highly concentrated $\rho \gtrsim \rho^{**}$ solutions and melts were not considered in the current study, but we can say that in this regime one must abandon the Brownian equation and use a second order equation of motion, such as the popular Dissipative Particle Dynamics\cite{espanol1995statistical} integrator. The pseudo-continuous $J\gg N$ model can still be applied similarly as in the present study, since the mechanistic chain model and the equation of motion employed to propagate that model in time are two separate things.

On passing, we emphasize that the linear spring interaction is appropriate for simulations of phenomena with a timescale $t\gg \tau$. As a counterexample, for an extreme shear flow $\dot{\gamma} \gtrsim \tau^{-1} = 10^6\, \text{s}^{-1}$ one will require more expensive FENE springs.

\section{The discrete model}
In this section we will provide a discrete counterpart to the continuous equation of motion, Equation~\eqref{brownian}, and integrate it over a short time step $\Delta t$. The details get a bit technical, but are worth following since a properly designed discretisation scheme is essential to suppress chain crossings.

The main idea is to sample the continuous backbone $s\in (0,1)$ using a finite number $j=1,\, 2,\, \ldots,\, J$ of discrete points as shown in Figure~\ref{model}. The potential of a fictitious $j$-``particle'' centered around $s_0 = (2j-1)/(2J)$ is
\begin{equation}\label{jpotential}
\Phi_0(\mathbf{r}-\mathbf{R}_{j}) = \int_{s_0-1/(2J)}^{s_0+1/(2J)} ds\, \Phi(\mathbf{r}-\mathbf{R}(s))  \approx \left(\frac{N}{J}\right) k_B T\exp \left(-\frac{(\mathbf{r}-\mathbf{R}_{j})^2}{2\lambda^2}\right).
\end{equation}

The choice $J=N$ corresponds to the simplest bead-and-spring model, which has gaps that allow chains to cross their paths. The choice $J=2N$ is similar to the situation obtained using SRP, except that in our case the potential on both the ``beads'' and the ``segments'' is exactly the same. In general, we will consider $J\gg N$ such that the largest gap $\texttt{max}|\mathbf{R}_j-\mathbf{R}_{j-1}|\ll \lambda$ is much smaller than $\lambda$. There exists a certain threshold, similar to the Nyquist rate in signal processing, beyond which the discrete model behaves just like the continuous Edwards chain would. We found that at equilibrium $J/N = 4$ is sufficient, whereas more points may be required in situations where the chains are stretched, such as under shear or in a polymer brush, or for more flexible chains with $\lambda < b$.

As is well known, the configuration of any given chain can equivalently be described by a set of Rouse\cite{rouse1953theory} modes  $\mathbf{a}_n = \int_0^1ds\, \mathbf{R}(s)\cos(\pi ns)$, where $n=0,1,2,\ldots, (N-1)$. In this work we retain $(N-1)$ modes $+1$ center of mass to be consistent with the number of blobs $N$. The equation of motion in the Rouse domain becomes
\begin{equation}\label{rouseeq}
\zeta \frac{\partial \mathbf{a}_n(t)}{\partial t} = -\left(\frac{3\pi^2 n^2 k_B T}{Nb^2}\right) \mathbf{a}_n(t) + \mathbf{\tilde{F}}_n(t) + \sqrt{(1+\delta_{0n})k_B T\zeta}\mathbf{\tilde{W}}_n(t)
\end{equation}
where the Wiener process is $\braket{\mathbf{\tilde{W}}_{cn}^{\alpha}(t) \mathbf{\tilde{W}}_{c'n'}^{\beta}(t')} = \delta^{\alpha \beta} \delta_{cc'}\delta_{nn'} \delta(t-t')$ and the spectral force
\begin{align}
\mathbf{\tilde{F}}_n &= -N\int_0^1 ds\, \cos (\pi n s) \nabla U(\mathbf{r})_{\mathbf{r}=\mathbf{R}(s)}\\
&\approx -\left(\frac{N}{J}\right)\sum_{j=1}^J \cos \left(\frac{\pi (2j-1)n}{2J}\right) \nabla U(\mathbf{r})_{\mathbf{r}=\mathbf{R}_{j}}\label{specforce}
\end{align}
is the discrete cosine transform of the real force. We must now integrate the continuous Rouse equation~\eqref{rouseeq} over a discrete time interval $\Delta t$. The main limitation on the time step is that two blobs repelling at maximum strength should not move further than their own size. This leads to $\Delta t \lesssim (6\pi \eta_s b^3)/k_B T = \tau$. However, when we integrate the random force over the same time step, the mean blob displacement is $\sqrt{\braket{\Delta \mathbf{R}^2}} = \sqrt{6k_B T \Delta t/(6\pi \eta_s b)} = \sqrt{6}b$. This distance is $\sqrt{6}\approx 2.4$ times greater than the coarse-grained excluded volume force range $\lambda = b$, and therefore would cause plenty of chain crossings. It is not surprising, since the concept of a ``blob'' (see Equation~\eqref{mapping}) was originally justified only in the thermodynamic long time $t\rightarrow \infty$ limit, while if observed at short times $t\approx \tau$ there are of course no such actual blobs. Therefore, to derive any meaningful information from our blob-based model, we must truncate the sampling rate of the random force. In particular, we propose to update the random force only once every $M\gg 1$ steps, while between the updates the blobs move ballistically with fixed increments of magnitude
\begin{equation}\label{rmagnitude}
\sqrt{\braket{\Delta \mathbf{R}^2}} = \sqrt{\frac{6 k_B T \Delta t}{(6\pi \eta_s b)M}}
\end{equation}
and random direction. In other words, we smooth out the Dirac delta correlation over a finite time span $(M\Delta t)$ while keeping the power spectrum at zero frequency unchanged, so that the long time properties are preserved but the instantaneous value of the force is smaller by $1/\sqrt{M}$. Specifically, the mean squared displacement over a long time $t\gg (M\Delta t)$ remains the same as in the continuum theory: $\braket{\Delta \mathbf{R}^2} = 6b^2 t/\tau$. We have used $M=120$, which gives a random displacement of $(\sqrt{6/120} \approx 0.22) b $ per step, sufficiently small to be repelled by the excluded volume force which pushes the two blobs apart by one $\lambda=b$ during the same time step $\Delta t$. A larger value of $M$ makes chain crossings less likely (see Figure~\ref{certainty}), at the expense of having to discard more short-time correlation data as unphysical, such as seen at short time scales in Figure~\ref{g3}.

The solution to Equation~\eqref{rouseeq} is written as
\begin{equation}\label{rousesolution}
\mathbf{a}_n (t+\Delta t) = \left\{ \mathbf{a}_n (t) + \frac{\Delta t}{\zeta} \mathbf{\tilde{F}}_n[\mathbf{a}_n(t)] + \sqrt{\frac{2 k_B T \Delta t}{\zeta M}}\mathcal{\tilde{R}}_n^{(3)}  \right\} \Big/ \left(1+\frac{\Delta t}{\tau_n}\right),
\end{equation}
where the spring relaxation times are
\begin{equation}\label{tm}
\tau_n = \frac{1}{3\pi^2}\left(\frac{6\pi \eta_s b^3}{k_B T}\right) \left(\frac{N}{n}\right)^2,
\end{equation}
and the symbol $\mathcal{\tilde{R}}_n^{(3)}$ stands for an isotropic random vector of mean zero and variance
\begin{equation}\label{fdt}
\braket{[\mathcal{\tilde{R}}_n^{(3)}]^2} = \frac{3}{2}(1+\delta_{0n}).
\end{equation}
To further minimize the largest possible displacement due to the random force, we use a uniform spherical distribution. First, generate $j=1,\,2,\ldots,\, N$ vectors $\mathcal{R}_j^{(3)}$ of fixed length $\sqrt{3}$ and random orientation. This corresponds to adding the random displacements directly on the beads shown in Figure~\ref{fig:1}. The spectral displacements are then obtained by
\begin{equation}\label{randseries}
\mathcal{\tilde{R}}_n^{(3)} = \frac{1}{\sqrt{N}}\sum_{j=1}^N \mathcal{R}_j^{(3)} \cos \left(\frac{\pi(2j-1)n}{2N}\right).
\end{equation}
One can verify that the variance is indeed
\begin{equation}
\braket{[\mathcal{\tilde{R}}_n^{(3)}]^2} = \frac{3}{N}\sum_{j=1}^N \cos^2 \left(\frac{\pi(2j-1)n}{2N}\right) = \frac{3}{2}(1+\delta_{0n})
\end{equation}
as imposed by Equation~\eqref{fdt}. It must be clear that the random force described above only makes sense in the limit of many steps $t\gg (M\Delta t)$. Our model does not contain any sensible microscopic information on the scale of a single step $t\lesssim (M\Delta t)$, where it cannot and should not be mapped to any real system.

\section{Computation of the excluded volume force}\label{force}
\begin{figure}[bht]
\centering
    \includegraphics[width=0.4\linewidth]{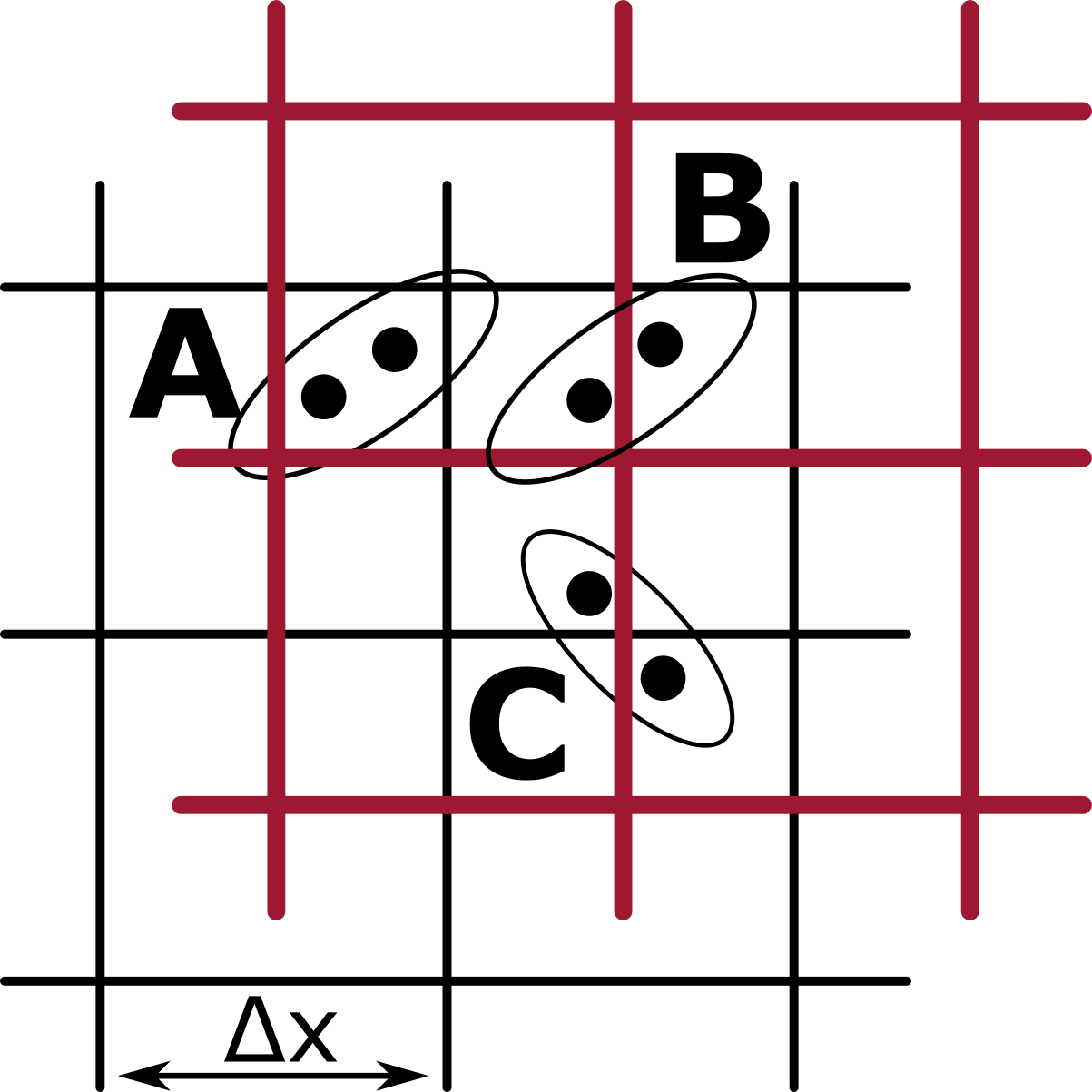}
    \caption{Interacting particles as seen by two staggered grids. Line color and thickness serve only as a visual aid.}\label{twogrids}
\end{figure}
The most time demanding step of the program is the calculation of the field gradient $\mathbf{F} = -\nabla U(\mathbf{r})$ at the position of every $j$-particle. The use of standard domain decomposition techniques\cite{plimpton1995fast} would require an execution time proportional to the total number of particles $CJ$ times the number of neighbors that each particle has, $\text{const.}\times J/N$, in total $O(CN(J/N)^2)$, which is a factor $(J/N)^2$ higher than a corresponding bead-and-spring simulation. Here we propose a mesh-based approximation which only takes $O(CN\log(CN))+O(CJ)$ computer time and does not suffer a significant slowdown in the important regime $J/N\gg 1$. We will need two rectangular grids, each having a large mesh size $\Delta x \lesssim \lambda$, and the origin staggered along all axes by half a spacing $\Delta x/2$ with respect to each other, as shown in Figure~\ref{twogrids}. The force on every particle $\mathbf{F}(\mathbf{R}_{c,j})$ is evaluated twice, using each of the two grids, and the average is fed to the equation of motion. We take into account the ``short range'' and the ``long range'' contributions. All particle pairs which share the same cell, such as the pair A, will interact via a short range routine. The pairs which fall into separate cells, such as the pair C, will interact via a long range routine. Lastly, borderline pairs such as B will interact via short range in one of the grids, and via long range in the other grid.

The first step in the force routine is to bin the coordinate $\mathbf{R}$ of each $j$-particle into its nearest cell in the central box:
\begin{equation}\label{periodic}
\begin{pmatrix}
k_x\\ k_y\\ k_z
\end{pmatrix}
= \texttt{ceil} \left(\left[\mathbf{R} - V^{1/3} \texttt{floor} \left( \frac{\mathbf{R}}{V^{1/3}}\right)\right]/\Delta x\right)
\end{equation}
where each 3D-cell index is $k_{\alpha} = 1,\, 2,\,  \ldots,\, K$, with $K=\texttt{round}\left(V^{1/3}/\Delta x\right)$ the total number of cells per spatial dimension, and $\Delta x = V^{1/3}/K$ re-adjusted so that $K$ is always an integer.

The short range routine is based on the linearization of the Gaussian force $\mathbf{F} = \mathbf{r}e^{-\mathbf{r}^2/(2\lambda^2)} \approx \mathbf{r}$ valid for $r\lesssim \lambda$. That way, the force on the particle located at $\mathbf{R}_p$ due to all the other nearby $Q$ particles which are in the same cell, is
\begin{equation}\label{fshort}
\mathbf{F}_{\text{short}}(\mathbf{R}_p) = \sum_{q=1}^Q (\mathbf{R}_p - \mathbf{R}_q) = Q\mathbf{R}_p - \sum_{q=1}^Q \mathbf{R}_q.
\end{equation}
The computational task is to count the total number $Q$ of particles in each cell, and sum all their coordinates $\sum \mathbf{R}_q$, followed by the cheap algebra of Equation~\eqref{fshort}, which costs little more than the very cheapest step in the code, Equation~\eqref{periodic}.

The interactions of particles across cell boundaries are taken into account by the long range routine. It is accurate for separations $r\gtrsim \Delta x$ and is the standard particle-mesh calculation\cite{cerutti2009staggered} which has been used for soft Gaussian potentials before\cite{baeurle2002field, zhang2013new}. Here we recycle the particle count $Q$ to reshape it into a three dimensional array $\rho(\mathbf{r}')$ and imagine that the particles are all located at the center $\mathbf{r}'=\left(k_x\mathbf{\hat{x}} + k_y\mathbf{\hat{y}} + k_z\mathbf{\hat{z}}\right)\Delta x$ of their corresponding cell. The force on every particle in a given cell $\mathbf{r}$ is then obtained by the convolution theorem:
\begin{equation}\label{flong}
\mathbf{F}_{\text{long}}(\mathbf{r}) = \texttt{IFFT} \left\{\texttt{FFT}[\rho(\mathbf{r}')]\cdot \texttt{FFT} \left[\mathbf{r}'e^{-\mathbf{r}'^2/(2\lambda^2)}\right]\right\}
\end{equation}
where $\texttt{(I)FFT}$ is the standard (Inverse) Fast Fourier Transform in three dimensions, which automatically incorporates the periodic boundary conditions. The total force on each particle is the sum
\begin{equation}\label{fgrid}
\mathbf{F}_{\text{total}} = \braket{\mathbf{F}_{\text{short}} + \mathbf{F}_{\text{long}}}_{\text{grid}},
\end{equation}
averaged over the two grids. 

The error suffered by this algorithm is eventually smeared over the redundant $j$-particles and the final spectral force in Equation~\eqref{specforce} is more trustworthy than it may seem judging from the real space perspective. We also wish to draw attention to the fact that even the most accurate evaluation of the interparticle force is only exact at one particular instant in time $t$, after which it is inevitably subject to the bias of the time integrator, which is usually $o(\Delta t^2)$ accurate in itself. Consistent with these reasons, we found that the simulation results were virtually identical for all grid sizes $\Delta x \leq \lambda$, so we kept $\Delta x = \lambda$ for maximum speed.

\section{Simulation algorithm}
In this section we consolidate all our ideas into an algorithm which is the basis for the computer code. The goal is to start with a configuration $\mathbf{a}_n(t)$ as the input and predict a thermodynamically likely future configuration $\mathbf{a}_n(t+\Delta t)$ as the output.
\begin{enumerate}
\item \textbf{Generate the $C\times N$ random vectors $\mathcal{R}_{c,n}$} of unit length $\sqrt{3}$ and random orientation. This step is updated only once every $M=120$ iterations.

\item \textbf{Sample the chain conformation in real space} using $J$ points indexed at regular intervals along the backbone $s$:
\begin{equation}\label{DCT3}
\mathbf{R}_j = \mathbf{a}_0 + 2\sum_{n=1}^{N-1} \mathbf{a}_n \cos\left(\frac{\pi(2j-1)n}{2J}\right)
\end{equation}
with $j=1,\, 2,\, \ldots,\, J$. These locations will be used to compute the excluded volume interaction between different chains. The complexity of this step is $O(J\log J)$ per chain, if evaluated using an efficient \texttt{FFT}-based code\cite{makhoul}.

\item \textbf{Evaluate the excluded volume force} $\mathbf{F}_j = -\nabla U(\mathbf{r})_{\mathbf{r}=\mathbf{R}_j}$ on each $j$-particle using the approximate Equation~\eqref{fgrid}. Then, convert it to the Rouse domain $\tilde{\mathbf{F}}_n$ using Equation~\eqref{specforce}.

\item \textbf{Integrate the equation of motion} using the Backwards Euler formula in Equation~\eqref{rousesolution} to obtain the new configuration $\mathbf{a}_n(t+\Delta t)$ which now includes the random walk, the excluded volume and the spring forces.

\item Repeat steps 2-4 for $M=120$ iterations using the same set of random displacements. Then, start over from step 1.

\item As a final remark, we note that the random numbers $\mathcal{R}_{c,j}$ do not instantaneously add up to zero, which leads to an overall diffusion of the entire system. Hence, we manually reset the system center of mass by translating all the particles
\begin{equation}
\mathbf{R}_{c,j}(t) \rightarrow \mathbf{R}_{c,j}(t) - \frac{1}{CJ}\sum_{c', j'}^{C,J}\mathbf{R}_{c',j'}(t)
\end{equation}
which guarantees $\sum \mathbf{R}_{c,j} = \text{const.} = 0$ at all times. This correction is required to remove the finite-size artifact from the trajectories, as explained in the appendix of reference \cite{kremer1990}.
\end{enumerate}

Before the start of the simulation, we need to decide on all the input parameters. As an example, suppose that we want to simulate polystyrene of molecular weight $M_w$ dissolved in toluene at density $\rho^{*}\ll \rho \ll \rho^{**}$. Using the mapping in Equation~\eqref{mapping} we convert this into the number of blobs $N$ and the blob size $\lambda$. \hl{Depending on the chemical species and temperature, one then has to choose the stiffness $\lambda/b$ and the excluded volume $v/\lambda^3$ parameters. In principle, any (positive) values are possible, but it will be computationally fastest to reach entanglement dynamics when both of these ratios are set equal to one, which is what we have done in the present study, and what seems to apply fairly well for a common system like polystyrene-toluene.}

Solvent viscosity, blob hydrodynamic radius $\approx b$ and temperature all coalesce to define the time unit $\tau = 6\pi \eta_s b^3 /(k_B T)$, but its absolute value is not important from the algorithm point of view, just like the absolute length $\lambda$ is not important, only the ratios $\lambda/b$ and $v/\lambda^3$.

\hl{Next, we need to impose either the pressure or the blob density of our system.} According to the semi-dilute theory, a polymer solution can be viewed as a melt of closely-packed correlation blobs, which leads to the simulation box size
\begin{equation}
V = \bar{v}_0\lambda^3 NC,
\end{equation}
where $\bar{v}_0$ is the dimensionless volume associated with a single blob. If $\bar{v}_0$ is too small, the blobs are too crowded and the interblob potential, Equation~\eqref{gaussian}, is unable to prevent chain crossings. If $\bar{v}_0$ is too big, the entanglement length grows and one needs longer chains to see the same level of chain interpenetration. We have found that a suitable compromise is $\bar{v}_0 = 2(4\pi/3)$. In terms of chains per unit volume,
\begin{equation}
\frac{C}{V} = \frac{1}{\bar{v}_0 \lambda^3 N} \propto \frac{\rho}{M_w}.
\end{equation}
This chain density ensures that the osmotic pressure scales as
\begin{equation}
\Pi \approx \frac{k_B T}{\lambda^3} \propto \rho^{9/4}
\end{equation}
which is the well-known des Cloiseaux law, and in our case it means that the pressure is the same regardless of chain length $N$. We have verified numerically that this is true for sufficiently long $N>32$ chains. Alternatively, one could fix the pressure $\Pi$ and let the box volume $V$ fluctuate in an isobaric fashion, but we have not tried this.

Finally, there are some technical/discretisation/finite-size settings: the number of chains $C/\sqrt{N} \gg 1$, the level of chain continuity \hl{$J/N \gg b/\lambda$}, and the grid size $\Delta x/\lambda \ll 1$. \hl{As for the time step, we must ensure that the excluded volume force does not overshoot its own range:}
\begin{equation}
\Delta R = \left(\frac{v}{\lambda^3}\right) \left(\frac{k_B T}{6\pi \eta_s b}\right) \left(\frac{\Delta t}{\lambda}\right) \ll \lambda
\end{equation}
\hl{which leads to time step limitation}
\begin{equation}
\frac{\Delta t}{\tau} \ll \left(\frac{\lambda}{b}\right)^2 \left(\frac{\lambda^3}{v}\right).
\end{equation}
\hl{Lastly, it is crucial that the random displacement be smaller than the repulsive one:}
\begin{equation}
b\sqrt{\frac{\Delta t}{M\tau}} \ll \Delta R = \left(\frac{v}{\lambda^3}\right) \left(\frac{k_B T}{6\pi \eta_s b}\right) \left(\frac{\Delta t}{\lambda}\right),
\end{equation}
\hl{which dictates the random force sampling cutoff:}
\begin{equation}
M \gg \left[\left(\frac{\lambda^3}{v}\right)\left(\frac{\lambda}{b}\right)\right]^2 \frac{\tau}{\Delta t}
\end{equation}
\hl{and gives the absolute shortest time scale beyond which the blob model is not applicable:}
\begin{equation}
t_{\text{allowed}} \gg t_{\text{min}} = M\Delta t = \tau \left[\left(\frac{\lambda^3}{v}\right)\left(\frac{\lambda}{b}\right)\right]^2.
\end{equation}
In the limits quoted above, our numerical algorithm is expected to approach the exact analytical solution for the multi-chain problem, Equation~\eqref{brownian}. Of course, computational time becomes very long, so initially we simulate the system with a reasonable choice $\Delta x = \lambda$ and calculate some physical property such as the diffusion coefficient. Then, we repeat the simulation with $\Delta x = 0.5\lambda$ and obtain an identical result, whereas $\Delta x = 2\lambda$ produces a significantly different outcome, and so we conclude that $\Delta x = \lambda$ is the upper safety limit. This test is repeated for all the technical parameters to ensure that the physical results do not depend on their choice.

The initially chosen configuration $\mathbf{a}_n (t = 0)$ should be close to thermal equilibrium which is a priori not known. To reach the equilibrium state quickly, we use Ref.~\cite{subramanian2010topology} method where every simulation is started with only $N=1$ blob per chain which is just Gaussian particles in a box, a model for a solution at density $\rho = \rho^*$, or the border between dilute and semi-dilute. After a few dozen iterations, the particles have repelled each other sufficiently and we can add the second mode $N=2$ to replace each ball with a randomly oriented Gaussian rod. After the rods have settled into their equilibrium distribution, we double the chain length again to $N=4$ and this process continues until the desired $N$ is obtained. The acquisition phase then starts where we record chain trajectories for analysis of various quantities and correlations of interest. We must acquire enough time steps to cover the dynamics for a time frame much longer than the system's own longest relaxation time.

\section{Results}

\begin{figure}[bht]
\centering
\begin{minipage}{.32\textwidth}
	\centering
		\begingroup
			\sbox0{\includegraphics{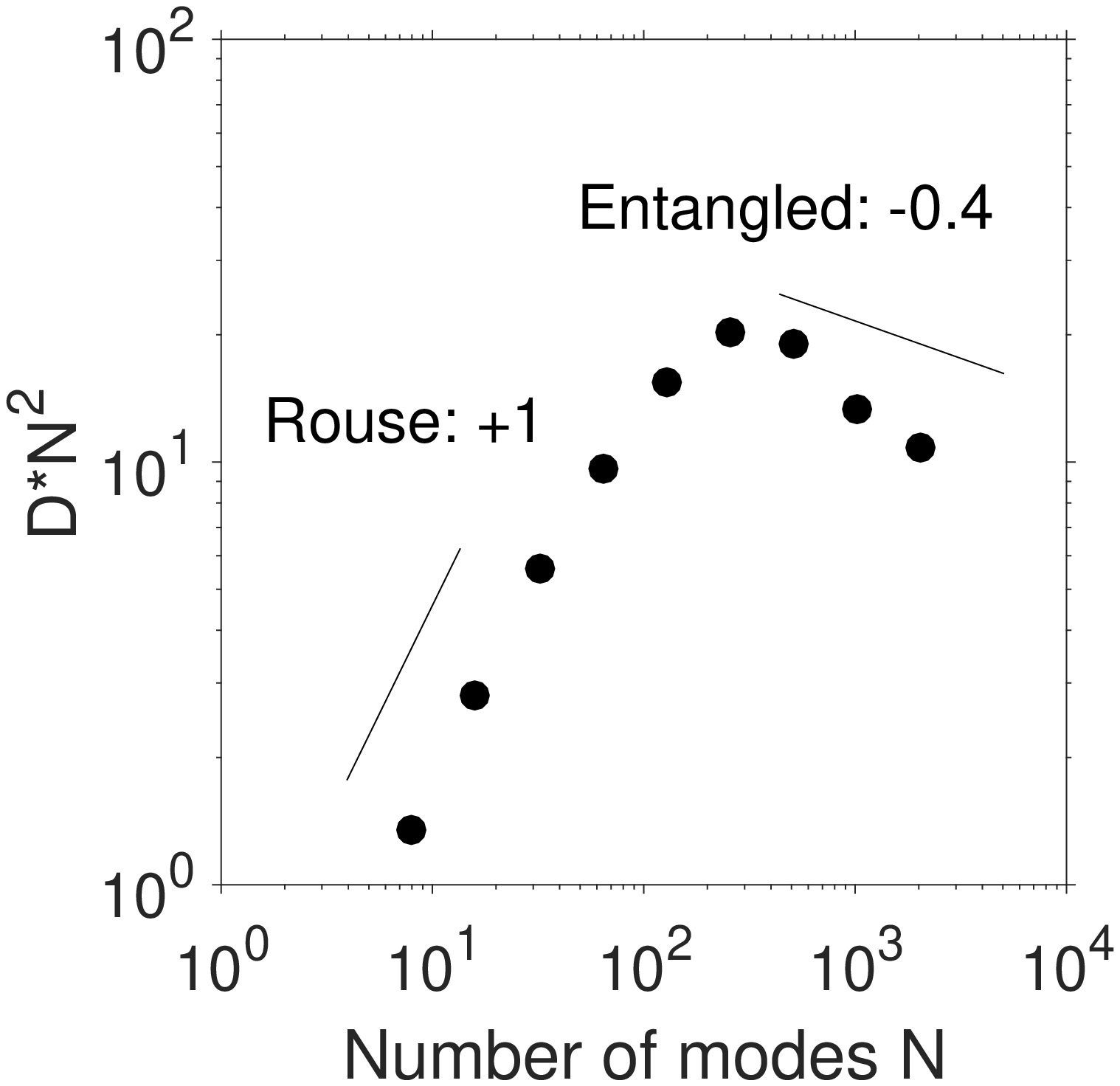}} % measure the original image size
			\includegraphics[clip,trim={.15\wd0} 0 {.2\wd0} 0,width=\linewidth]{./fig/DN.eps} % crop 15% on the left and 20% on the right
		\endgroup
    \caption{Self-diffusion $DN^2$}
    \label{DN}
\end{minipage}
\hfill
\begin{minipage}{.32\textwidth}
	\centering
		\begingroup
			\sbox0{\includegraphics{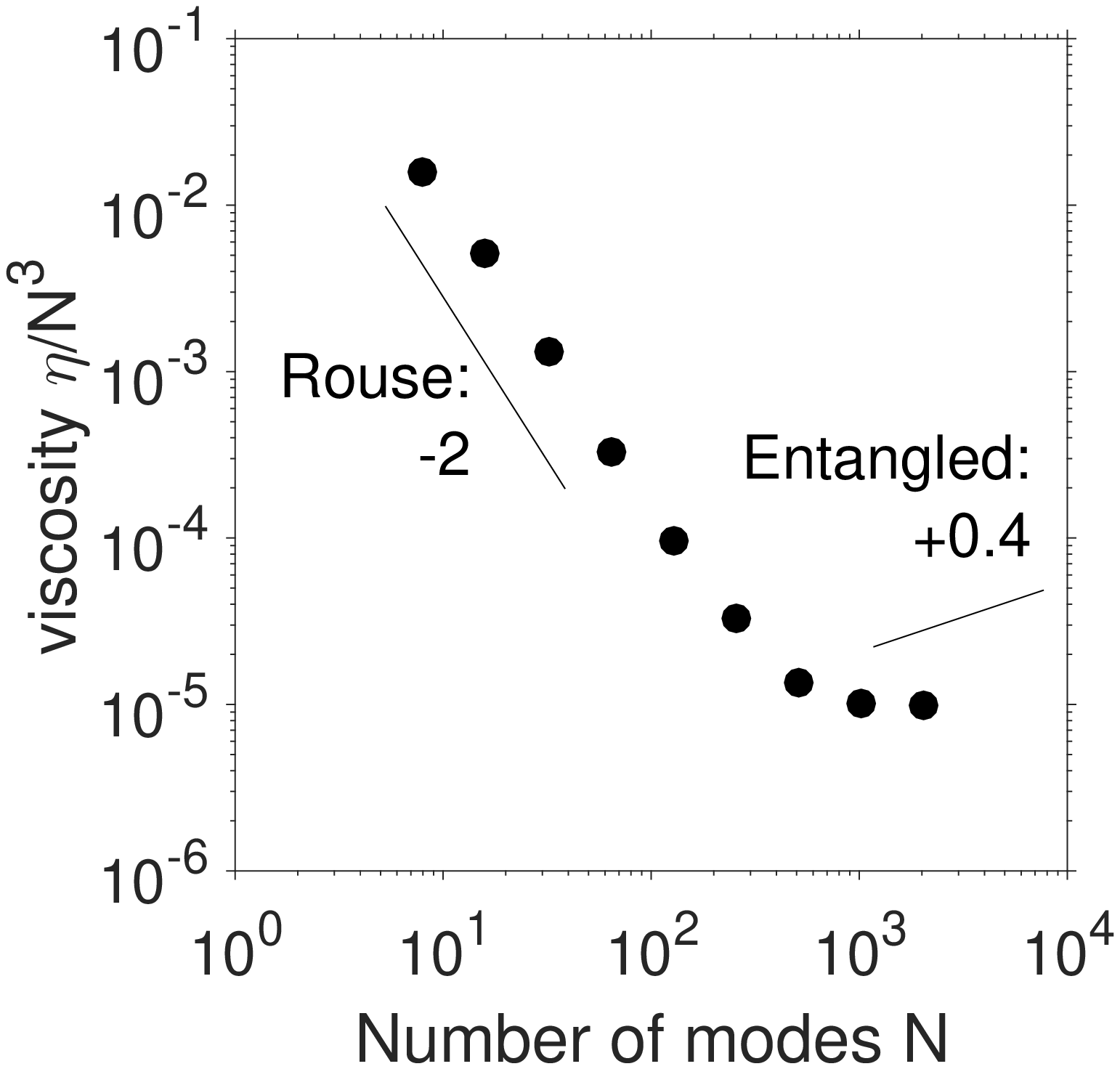}} % measure the original image size
			\includegraphics[clip,trim={.15\wd0} 0 {.2\wd0} 0,width=\linewidth]{./fig/etaN.eps} % crop 15% on the left and 20% on the right
		\endgroup
    \caption{Viscosity $\eta/N^3$}
    \label{etaN}
\end{minipage}
\hfill
\begin{minipage}{.32\textwidth}
	\centering
		\begingroup
			\sbox0{\includegraphics{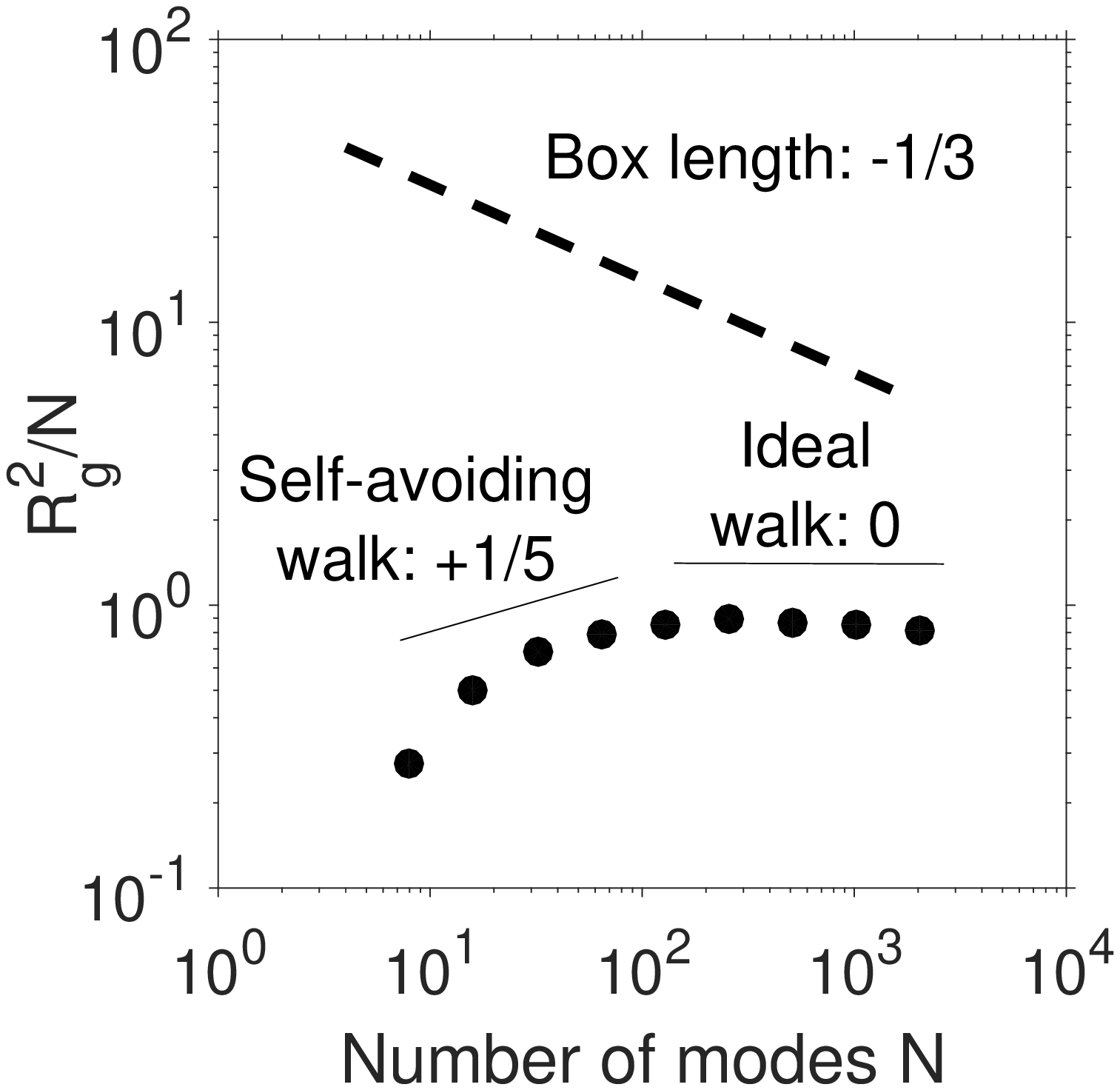}} % measure the original image size
			\includegraphics[clip,trim={.15\wd0} 0 {.2\wd0} 0,width=\linewidth]{./fig/Rg.eps} % crop 15% on the left and 20% on the right
		\endgroup
    \caption{Radius $R_g^2/N$}
    \label{Rg}
\end{minipage}
\end{figure}

The purpose of this section is to demonstrate the feasibility and usefulness of our newly developed simulation method. A reasonably fast implementation was achieved by writing a custom MATLAB executable subroutine containing CUDA code and running on an Nvidia Quadro M4000 GPU. The computation time was about $4\times 10^{-7}\,\text{s}$ per time step, per chain, per Rouse mode. The source code is available upon request to the corresponding author. 
 
We have simulated $C=64$ chains with the number of Rouse modes spanning $N=8,\,16,\,32,\,64,\,128,\,256,\,512,\,1024,\, \text{and}\, 2048$, while keeping all other parameters constant. The primitive path analysis of an equilibrated static configuration for the $N=1024$ chains was performed using the Z1 code\cite{karayiannis2009combined} available online, which found $Z=17.5$ entanglements per chain. The entanglement length is thus $N_e = N/Z = 59$, quite consistent with the departure from Rouse dynamics seen in Figures~\ref{DN} and \ref{etaN}.

The longest run with $N=2048$ modes lasted for about four months and was enough to reach one relaxation time as can be seen from the emerging plateaus in Figures~\ref{g3} and \ref{eta}. This computational effort was sufficient to clearly reveal the departure from Rouse dynamics and into the entangled regime.

\subsection{Mean squared displacement}
\begin{figure}[bht]
  \begin{subfigure}[b]{0.49\textwidth}
	  \begingroup
			\sbox0{\includegraphics{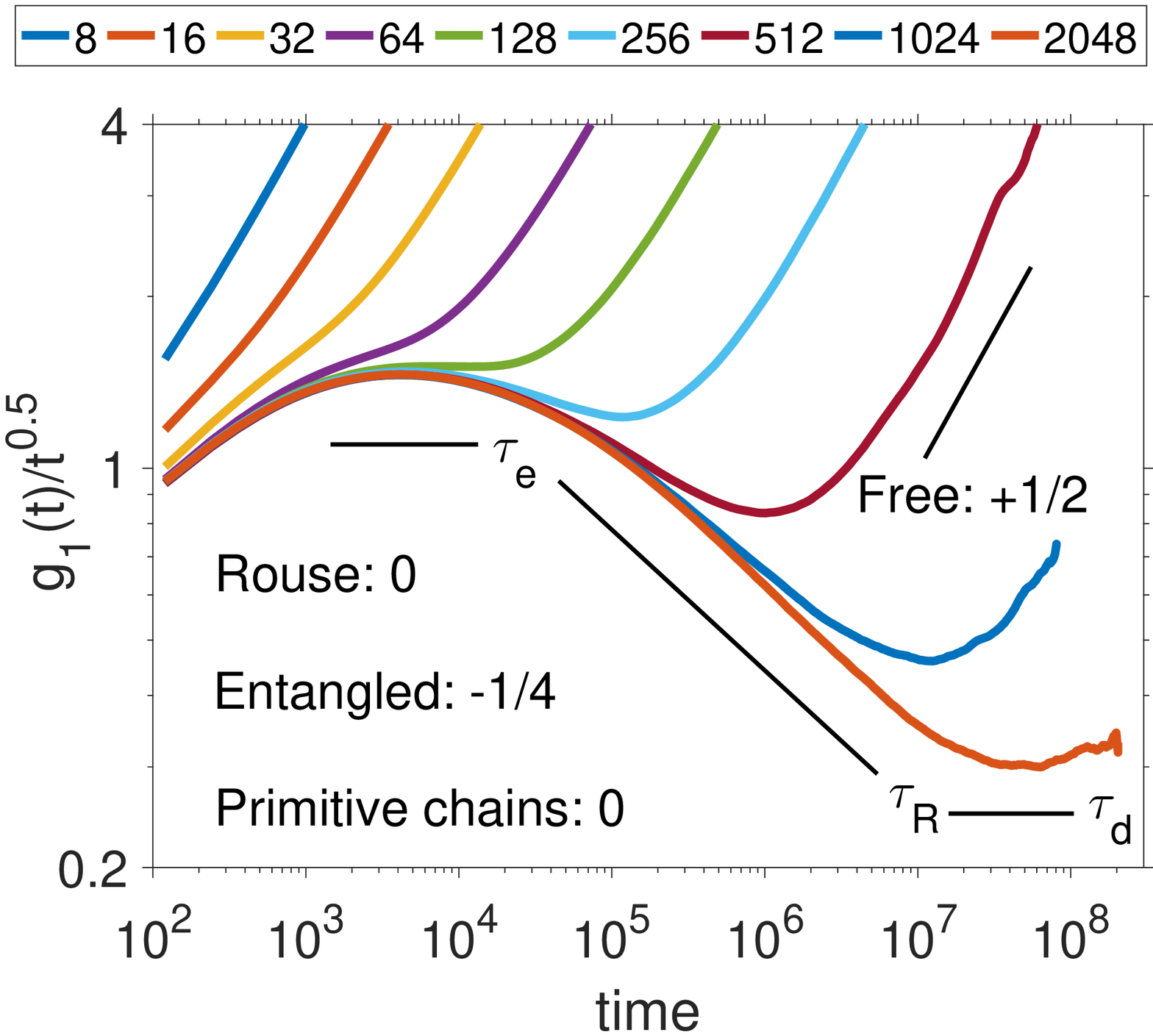}} % measure the original image size
			\includegraphics[clip,trim={.08\wd0} 0 {.1\wd0} 0,width=\linewidth]{./fig/g1.eps} % crop 15% on the left and 20% on the right
		\endgroup
    \caption{Central monomer MSQD divided by $\sqrt{t}$}
    \label{g1}
  \end{subfigure}
  \begin{subfigure}[b]{0.49\textwidth}
	  \begingroup
			\sbox0{\includegraphics{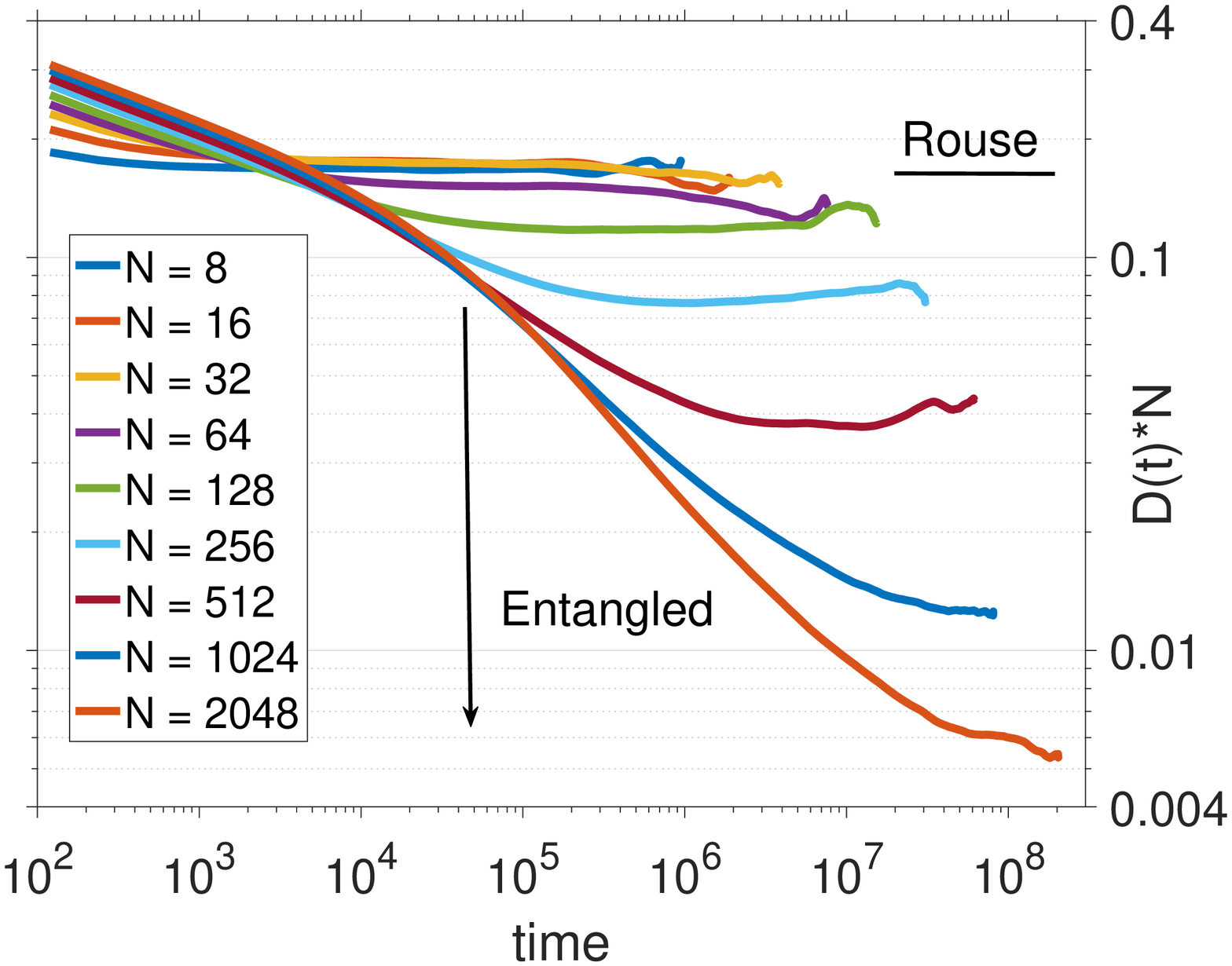}} % measure the original image size
			\includegraphics[clip,trim={.1\wd0} 0 0 0,width=\linewidth]{./fig/g3.eps} % crop 15% on the left and 20% on the right
		\endgroup
    \caption{Center of mass MSQD multiplied by $N/6t$}
    \label{g3}
  \end{subfigure}
	\caption{Mean squared displacement}\label{simd}
\end{figure}
Perhaps the most famous fact about entangled polymers\cite{de1979scaling} is that their motion is confined to an imaginary tube, created by the constraints imposed by all other nearby chains. The strongest topological constraint is felt by the central $j=J/2$ monomer, whereas the chain ends $j=1,\,J$ are more mobile and show less reptation. It is rather well established that the mean squared displacement of the central monomer
\begin{equation}
g_1(t) = \braket{[\mathbf{R}_{J/2}(t) - \mathbf{R}_{J/2}(0)]^2}
\end{equation}
scales as $g_1 \propto t^{1/4}$ in the range $(\tau_e \approx \tau N_e^2)< t < (\tau_R \approx \tau N^2)$, which is a signature of anisotropic diffusion along the randomly curved tube, in the presence of chain countour length fluctuations. In contrast, unentangled phantom chains would scale as $g_1 \propto t^{1/2}$ at the slowest, as described by the Rouse model \hl{with full details available in a textbook reference}\cite{likhtman2012viscoelasticity}. Therefore, we plot $g_1(t)/\sqrt{t}$ in Figure~\ref{g1}, where the negative slope of $t^{-1/4}$ clearly indicates a departure from Rouse dynamics and the onset of reptation. For the very longest chains $N=2048$ we start to see the beginnings of a new dynamical regime $(\tau_R \approx \tau N^2) < t < (\tau_d \approx \tau (N/N_e)^3)$, where the contour length fluctuations die out and pure reptation starts to dominate: $g_1 \propto t^{1/2}$ once again.

From an experimentalist point of view, it is more common to measure the mean squared displacement of the center of mass,
\begin{equation}
g_3 = \braket{[\mathbf{a}_{0}(t) - \mathbf{a}_{0}(0)]^2},
\end{equation}
which can be used to calculate the self-diffusion coefficient
\begin{equation}
D = \lim_{t\rightarrow \infty} \frac{g_3(t)}{6t}.
\end{equation}
Phantom chains would scale as $D_1 = k_B T/(6\pi \eta_s b N)$ which is the result for a group of $N$ independent random walkers. However, entangled chains are confined to move in a tube of length $L \propto N\lambda$, and the time it takes to diffuse this far is $\tau_{\text{tube}} \propto L^2/D_1 \propto N^3$. During this time the chain center of mass has been displaced a distance of about its own radius of gyration $R_g \propto N^{1/2} \lambda$, so the actual self-diffusion coefficient is $D \propto R_g^2/\tau_{\text{tube}} \propto k_B T/(6\pi \eta_s b) N^{-2}$.

To emphasize the cross-over from Rouse to entanglement, we plot $g_3 N/6t$ in Figure~\ref{g3}. In the long time $t\rightarrow \infty$ limit a plateau develops and its value gives the self-diffusion coefficient $DN$. On the short-time scale one expects all $g_3 N/6t$ curves to overlap, but instead we see a slight overshoot due to the way that the random force was implemented (Equation~\eqref{rmagnitude}) in our equation of motion. Therefore, our data can only be considered valid after at many $M$-sized random steps, which is to say $t\gg (M\Delta t) = 120\tau$.

To focus on the entanglement behaviour, the result is rescaled to $DN^2$ and plotted in Figure~\ref{DN}. The shape of $D(N)$ is quite similar to the one measured in experiments\cite{leger1981reptation, callaghan1981self} and other simulations\cite{putz2000entanglement}, and specifically it takes about one decade worth of $N$ to transit from $D=1/N$ to $D=1/N^2$ slope. We have not made a direct comparison to experiment in order to avoid a bias in the design of our own algorithm. As a side note, we mention that more elaborate theories\cite{frischknecht2000} invoke contour length fluctuation in addition to pure reptation, and predict $D\propto N^{-2.4}$. This law is obeyed fairly well by our last three points $512<N<2048$.

\subsection{Shear relaxation modulus}
\begin{figure}[bht]
  \begin{subfigure}[bht]{0.45\textwidth}
		  \begingroup
			\sbox0{\includegraphics{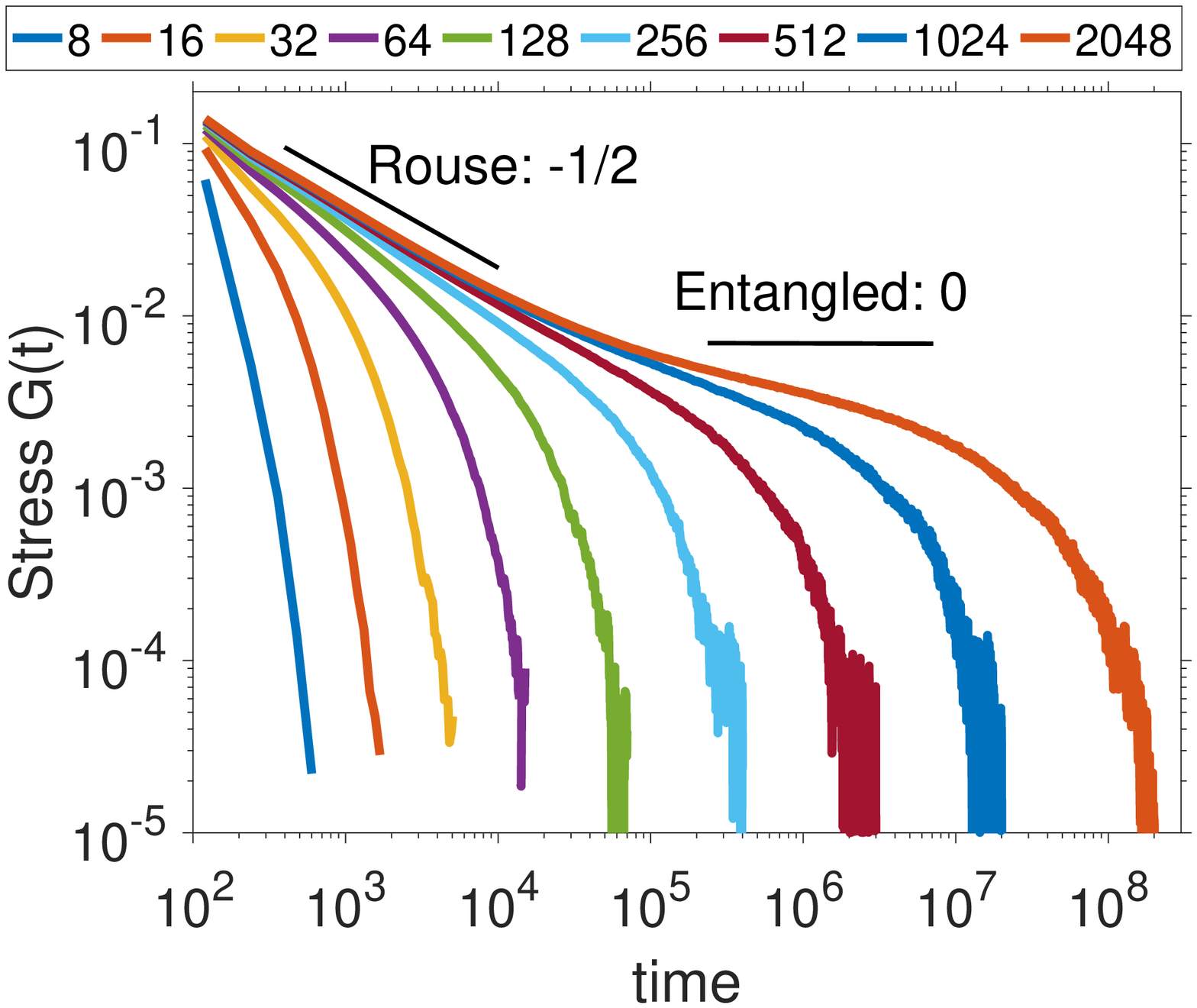}} % measure the original image size
			\includegraphics[clip,trim={.08\wd0} 0 {.15\wd0} 0,width=\linewidth]{./fig/Gt.eps} % crop 15% on the left and 20% on the right
		\endgroup
    \caption{Shear relaxation modulus $G(t) = \braket{\sigma_{xy} (t) \sigma_{xy}(0)}$ in units of $k_B T/b^3$}
    \label{Gt}
  \end{subfigure}
	\begin{subfigure}[bht]{0.5\textwidth}
		  \begingroup
			\sbox0{\includegraphics{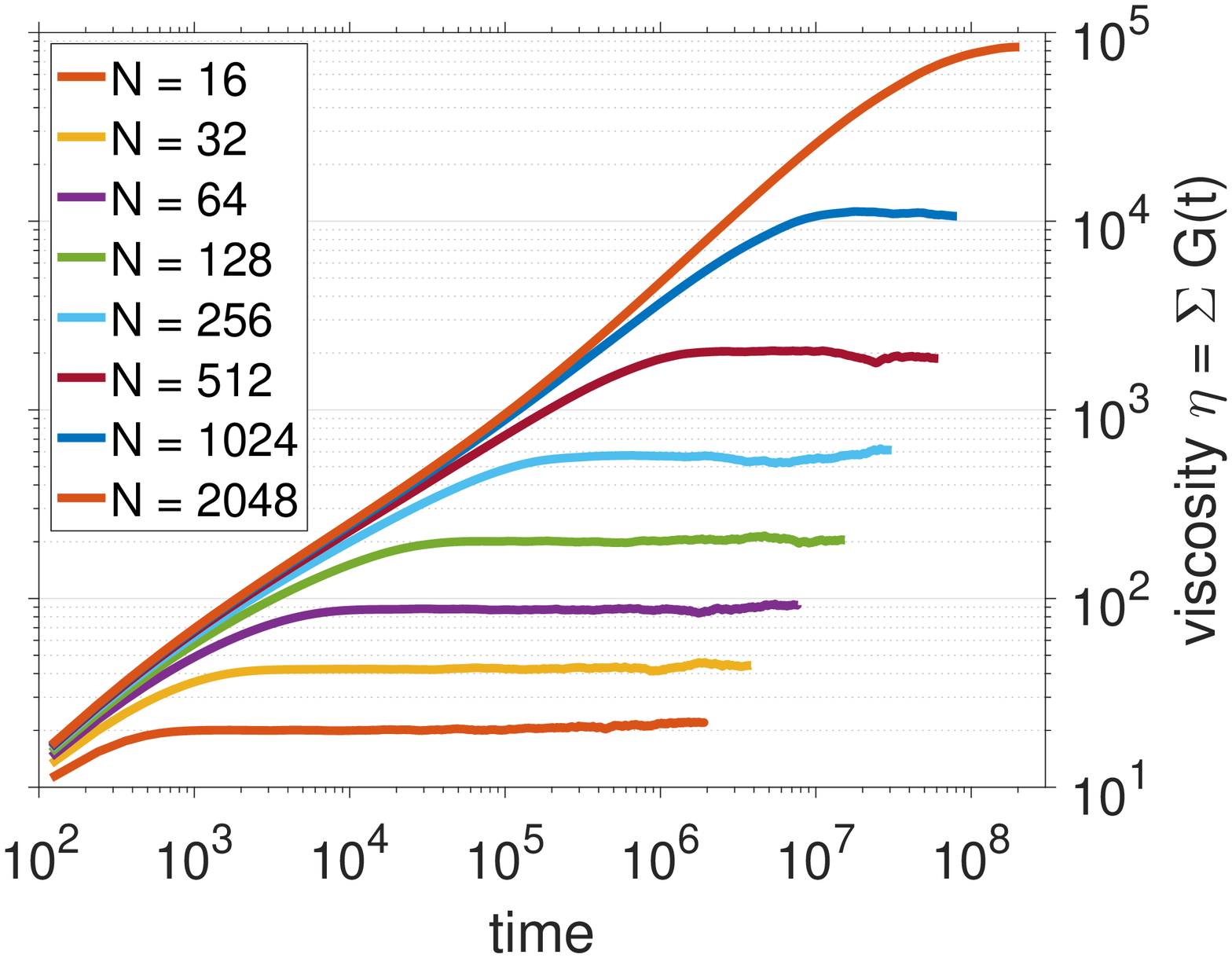}} % measure the original image size
			\includegraphics[clip,trim={.15\wd0} 0 {0.0\wd0} 0,width=\linewidth]{./fig/eta.eps} % crop 15% on the left and 20% on the right
		\endgroup
    \caption{Polymeric viscosity $\eta = \int G(t)\, dt$ in units of $6\pi \eta_s$}
    \label{eta}
  \end{subfigure}
  \begin{subfigure}[bht]{0.45\textwidth}
		  \begingroup
			\sbox0{\includegraphics{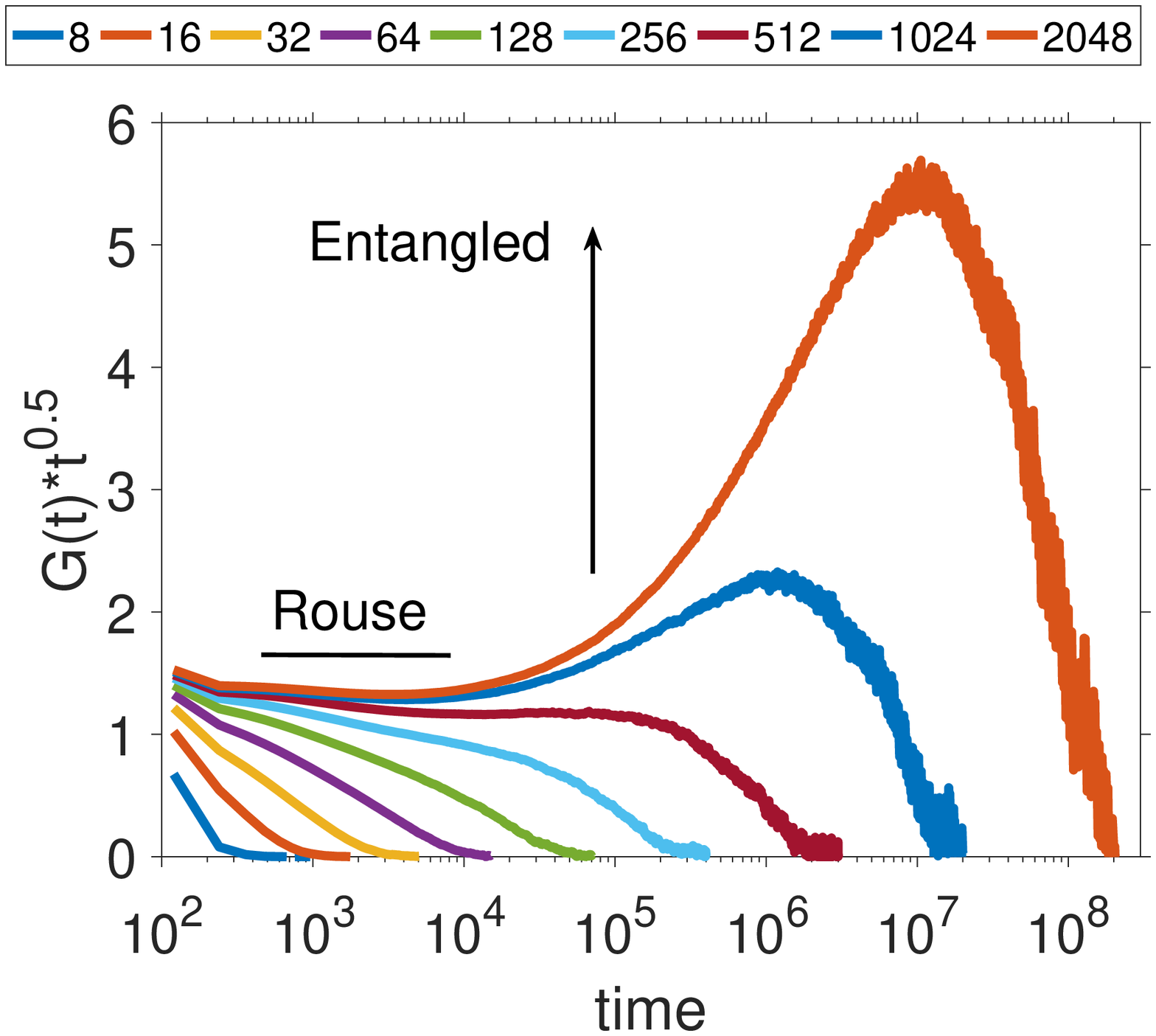}} % measure the original image size
			\includegraphics[clip,trim={.1\wd0} 0 {.15\wd0} 0,width=\linewidth]{./fig/Gtt.eps} % crop 15% on the left and 20% on the right
		\endgroup
    \caption{Rescaled $G(t)\sqrt{t}$}
    \label{Gtt}
  \end{subfigure}
	\caption{Analysis of shear stress fluctuations at equilibrium}\label{greenkubo}
\end{figure}

For long $N\gtrsim 100$ chains the instantaneous shear stress is dominated by the spring force contribution and is calculated\cite{doi1988theory} by
\begin{align}
\sigma^{\alpha \beta} &= -\frac{1}{V}\sum_{p=1}^P \mathbf{F}_p^{\alpha} \mathbf{R}_p^{\beta}\\
&= \frac{6\pi^2 k_B T}{VNb^2} \sum_{c=1}^C \sum_{n=1}^{N-1} n^2 \mathbf{a}_{c,n}^{\alpha} \mathbf{a}_{c,n}^{\beta}.
\end{align}
Various mechanical and rheological properties can be deduced from the knowledge of the shear stress autocorrelation function, also known as the shear relaxation modulus:
\begin{equation}
G(t) = \frac{V}{10 k_B T}\sum_{\alpha, \beta=1}^3 \braket{P^{\alpha \beta}(t) P^{\alpha \beta}(0)},
\end{equation}
where $P^{\alpha \beta}$ is the traceless portion of the stress tensor $\sigma^{\alpha \beta}$, defined as
\begin{equation}
P^{\alpha \beta} = \sigma^{\alpha \beta} - \frac{1}{3}\delta^{\alpha \beta} \sum_{\gamma = 1}^3 \sigma^{\gamma \gamma}.
\end{equation}
This formula\cite{mondello1997} utilizes all six stress components for best possible statistics. Further, we register the stress of each chain $\sigma^{\alpha \beta}_c$ separately and only include intrachain $c=c'$ correlations\cite{vladkov2006linear}, since the interchain contribution $\braket{\sigma_c^{\alpha \beta}(t) \sigma_{c'\neq c}^{\alpha \beta}(0)}$ should in theory average to zero in the long run, and therefore provides little valuable information, only useless noise.

The resulting $G(t)$ is plotted in Figure~\ref{Gt}. This can be compared to bead-and-spring simulations for melts, presented in Ref.~\cite{likhtman2007linear}. We can see that the overall number of time steps, about $10^8$, as well as the range of the $G(t)$ axis, about $10^4$, is similar in both types of simulations. The main difference is that our chains are based on soft blobs, which ultimately lead to unrealistic behaviour on short time scales, but the long time scales are reasonable and can be mapped to semi-dilute solutions described by $N \propto \rho^{5/4} M_w$ blobs of unlimited size $\lambda \propto \rho^{-3/4}$. In contrast, the KG model assumes very specific FENE-WCA interactions designed to reproduce short-time behavior in melts, which is then a strong limitation from the polymer solution point of view.

For entangled chains, either molten or semi-dilute, one expects a plateau $G_0$ to emerge with a value of roughly
\begin{equation}
G_0 = \nu k_B T,
\end{equation}
where $\nu$ is the number of entanglements per unit volume which in our case can be estimated as $\nu = ZC/V = 0.002/b^3$. Our chains are not sufficiently long yet to see an actual flat plateau, but judging from the trend in Fig.~\ref{Gt}, in particular the inflection point where the $G(t)$ slope starts to be flatter than $t^{-0.5}$, a value between $0.001$ and $0.005$ seems quite reasonable.

Further, we can estimate the zero-shear viscosity using the Green-Kubo relationship: $\eta(t) = \int_0^t G(t')\,dt' \approx \Delta t \sum G(t')$, plotted in Figure~\ref{eta}. The actual viscosity is obtained in the limit of $t\rightarrow \infty$, so in practice we must simulate long enough for the integral to flatten out, which can then be extrapolated to infinity and its value recorded in Fig.~\ref{etaN}. A simple reptation argument predicts the viscosity $\eta \propto N^3$, but most experiments and detailed theories\cite{milner1998reptation} quote the law as $\eta \propto N^{3.4}$. We therefore rescale our data to $\eta/N^3$ and find that while our longest chains are clearly not Rouse anymore ($\eta \approx N^{3.0}$), they are unfortunately not yet long enough to exhibit the experimental law. This is not surprising and is in fact corroborated by other experiments and single-chain models\cite{likhtman2005single} which agree that the onset of fully entangled dynamics in terms of viscosity occurs at larger $N$, compared to the structural correlations such as self-diffusion (see Fig.~\ref{DN}). Other multi-chain bead-and-spring simulations\cite{kroger2000rheological} do report slopes exceeding $+3$, but they are extrapolated from data under shear flow. With our present model we have not yet performed such non-equilibrium simulations.

For phantom chains which can cross each other, the shear stress relaxation modulus should behave according to the Rouse model:
\begin{equation}
G^{\text{Rouse}} (t) \propto \frac{1}{N}\sum_{n=1}^N e^{-3\pi^2 (t/\tau) (n/N)^2} \approx \int_0^N \frac{dn}{N}\, e^{-3\pi^2 (t/\tau) (n/N)^2} \approx \sqrt{\frac{\tau}{t}}.
\end{equation}
At large $N$, a power-law decay emerges: $G(t) \propto t^{-0.5}$, valid for timescales $1 \ll 3\pi^2 t/\tau \ll N^2$. In contrast, the simulated data $G(t)\sqrt{t}$, plotted in Figure~\ref{Gtt}, shows that for chains $N=512$ and longer, our stress relaxation is clearly slower than $t^{-0.5}$. This is a further indication that we are entering the entangled regime.

\subsection{Radius of gyration}
The focus of this paper is on the dynamics of entangled polymers, but for the sake of completeness we also present one static quantity, namely the radius of gyration:
\begin{align}
R_g^2 &= \int_0^1 |\mathbf{R}(s)-\mathbf{a}_0|^2\, ds = 2\sum_{n=1}^{\infty} |\mathbf{a}_n|^2\\
& \approx \left(2\sum_{n=1}^{N-1} |\mathbf{a}_n|^2\right) + R_0^2,
\end{align}
where one may optionally add a constant $R_0 \approx \lambda$ to compensate for all the higher Rouse modes which were truncated. We have plotted $R_g^2/N$ in Figure~\ref{Rg} to show that for long chains, the scaling is $R_g \propto \sqrt{N}$, and therefore the excluded volume force is fully screened and the chains obey ideal random walk statistics.

We also show that the largest radius of gyration is roughly a factor of three smaller than the size of the box $V^{1/3}$, which should be enough to ensure that the chains do not interact with their own periodic selves.

\subsection{Test for chain crossings}
\begin{figure}[bht]
\centering
\begin{minipage}{.45\textwidth}
  \centering
 \includegraphics[width=\linewidth]{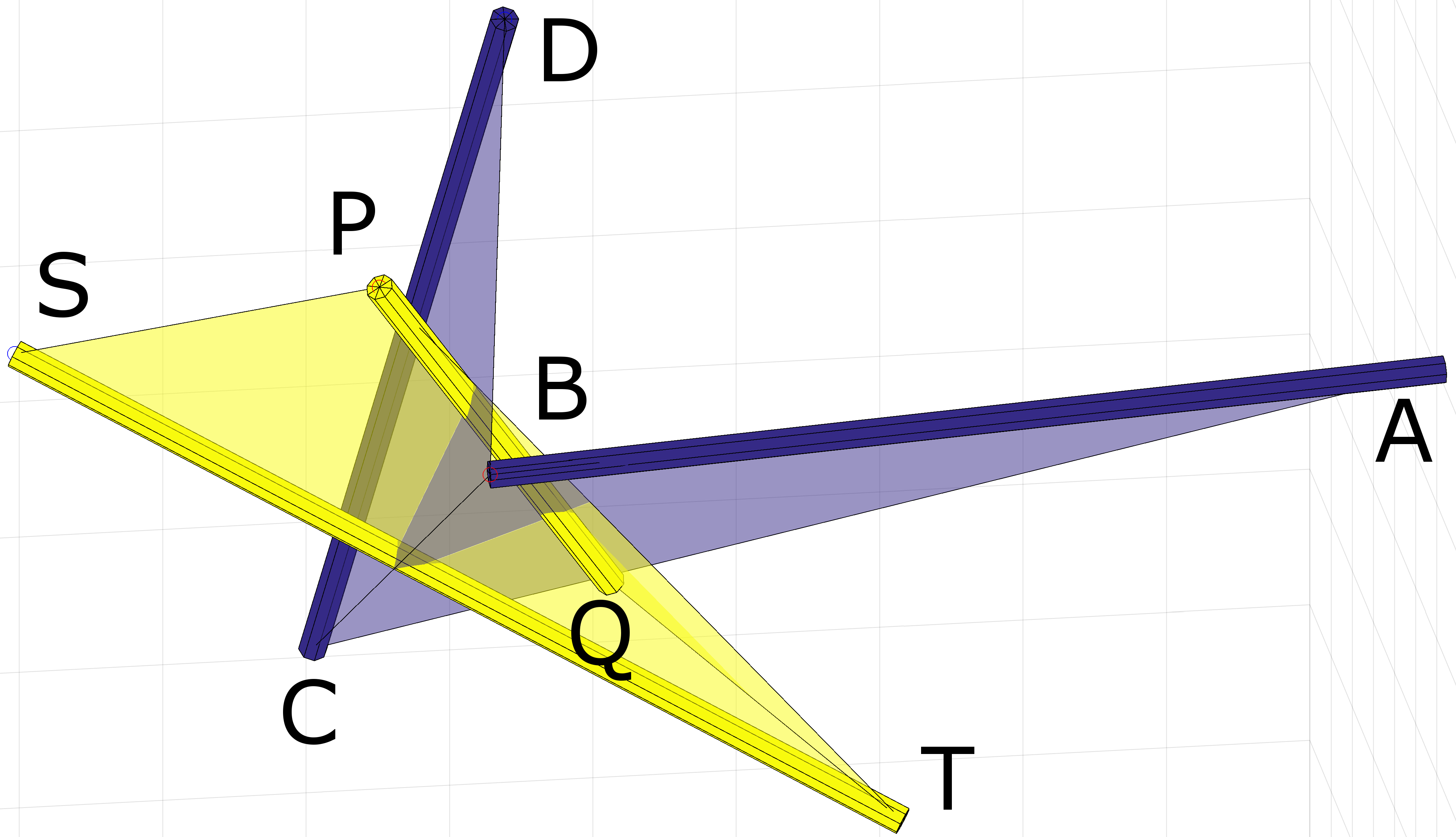}
    \caption{Topological analysis of two moving rods detects a crossing in this particular example}\label{rods}
\end{minipage}%
\hfill
\begin{minipage}{.45\textwidth}
  \centering
	
			\begingroup
			\sbox0{\includegraphics{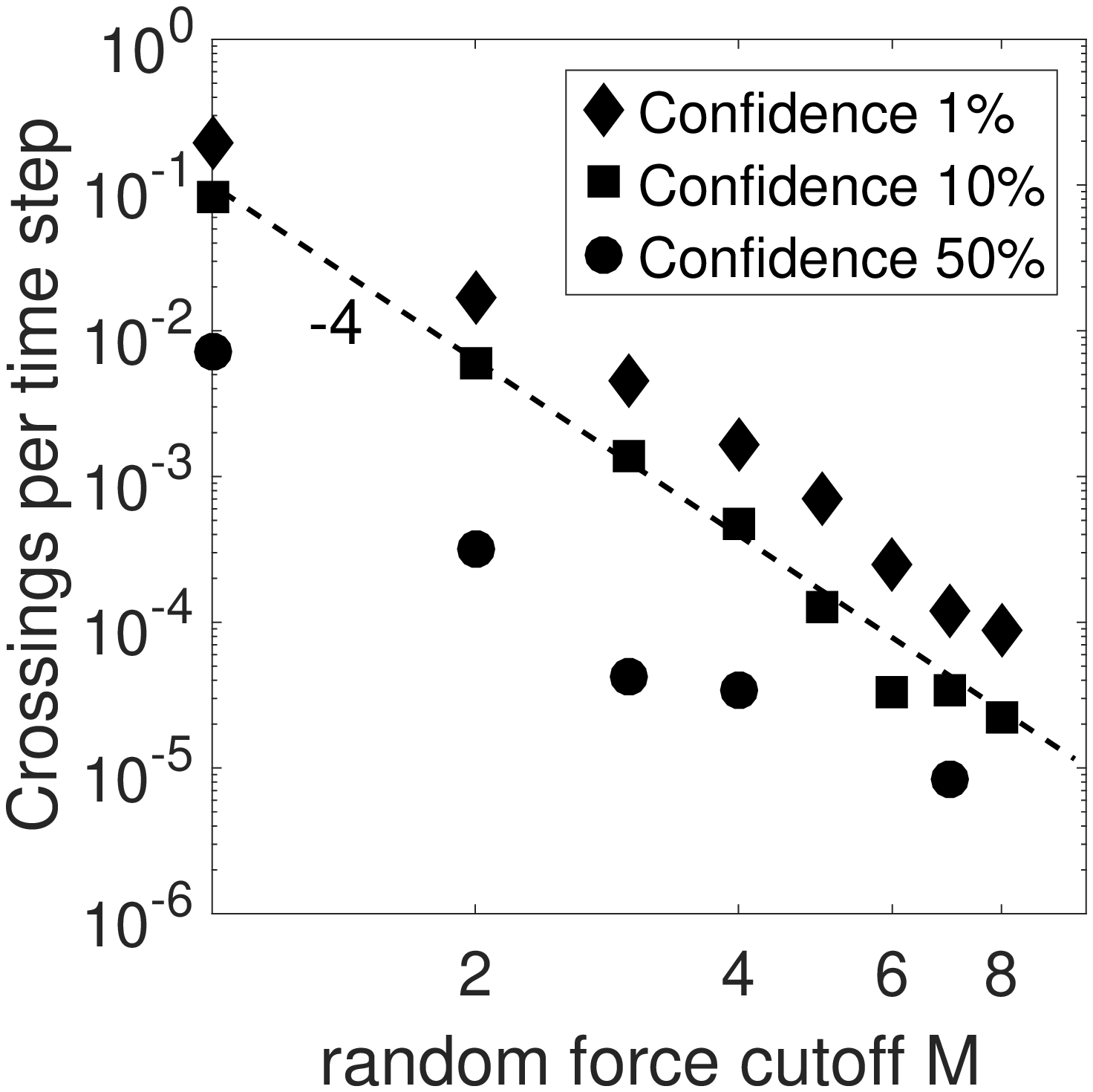}} % measure the original image size
			\includegraphics[clip,trim={.15\wd0} 0 {.2\wd0} 0,width=\linewidth]{./fig/crossings.eps} % crop 15% on the left and 20% on the right
		\endgroup
    \caption{Suspected crossing rate of $C=8$ chains and $N=16$ modes}\label{certainty}
\end{minipage}
\end{figure}

So far we have analyzed various physical properties of our simulation and they all indicate the emergence of reptation dynamics for long chains. This implies that chain crossings are unlikely on the scale exceeding the longest relaxation time $\tau_d \propto N^3$. To further strengthen the validity of our model, we now also present a direct calculation of chain crossing un-likelihood.

We shall analyze how the chain arrangement with respect to each other evolves from one time step to the next, and use a geometrical argument to estimate whether a crossing may have taken place. Every segment $(\mathbf{R}_j-\mathbf{R}_{j+1})(t) = \mathbf{AB}$ sweeps out a surface area, as it moves to its new position $(\mathbf{R}_j-\mathbf{R}_{j+1})(t+\Delta t) = \mathbf{CD}$. This unknown surface can be approximated by two adjoining triangles $\mathbf{ABC}$ and $\mathbf{BCD}$ (although a choice $\mathbf{ABD} \rightarrow \mathbf{ACD}$ is also possible and may produce a different result).

Before we can start the analysis, we need a mathematical criterion to determine if a fixed rod $\mathbf{PQ} = (\mathbf{P}-\mathbf{Q})$ intersects a fixed triangle $\mathbf{ABC}$. The rod $\mathbf{PQ}$ is defined by the set of all points
\begin{equation}
\mathbf{x}(t) = \mathbf{P} + (\mathbf{Q}-\mathbf{P})t, \quad (t>0)\, \&\, (t<1)
\end{equation}
(the parameter $t$ is not to be confused with the time variable), while the triangle $\mathbf{ABC}$ is the set of all points
\begin{align}
\begin{split}
\mathbf{y}(u,v) &= \mathbf{A} + (\mathbf{B}-\mathbf{A})u + (\mathbf{C}-\mathbf{A})v,\\
&{} (u>0)\, \&\, (v>0)\, \&\, (u+v)<1
\end{split}
\end{align}
To find the intersection $\mathbf{x}(t) = \mathbf{y}(u,v)$ we must solve the system of three equations
\begin{equation}
\mathbf{PQ}t + \mathbf{BA}u + \mathbf{CA}v = \mathbf{PA}
\end{equation}
and find the three unknowns
\begin{equation}
\begin{pmatrix}
t\\ u\\ v
\end{pmatrix} = \frac{1}{\mathbf{PQ}\cdot (\mathbf{BA}\times \mathbf{CA})} \begin{pmatrix}
\mathbf{PA}\cdot (\mathbf{BA}\times \mathbf{CA})\\
\mathbf{PA}\cdot (\mathbf{CA}\times \mathbf{PQ})\\
\mathbf{PA}\cdot (\mathbf{PQ}\times \mathbf{BA})
\end{pmatrix}
\end{equation}
Since the triangle is merely an approximation for the true (unknown) surface, we assign an intersection certainty score
\begin{equation}
I^{(1)} = f(t)f(u)f(v)f(1-t)f(1-u-v)
\end{equation}
where $f(x) = (1+\text{erf}(x/\sigma))/2$ is a fuzzy step function with a fuzziness parameter which we fixed to $\sigma = 0.05$, whereas a choice $\sigma \rightarrow 0$ would lead to an unrealistically crisp logic. A second test $I^{(2)}$ is performed with the same rod $\mathbf{PQ}$ and the second triangle $\mathbf{BCD}$. A score $I=1$ means that the rod clearly intersects the triangle, $I=0$ means that the rod is clearly outside the triangle, while some intermediate value $I\approx 0.5$ signals that the intersection is very close to the edge of the triangle and/or extremity of the rod, and the test result should be interpreted with caution.

As explained in Ref.~\cite{Holleran2007}, a fixed rod $\mathbf{PQ}$ crosses the path of another moving rod $\mathbf{AB}$ as it travels to $\mathbf{CD}$ if and only if one of the triangles $\mathbf{ABC}$ or $\mathbf{BCD}$ intersects the fixed rod $\mathbf{PQ}$, and the other triangle does not. If both or neither of the triangles intersect the rod, the crossing has not taken place. Logically, this is an ``exclusive or'' operation, which for a fuzzy input is computed\cite{hernandez2011} as
\begin{equation}
I = \texttt{xor}\left(I^{(1)}, I^{(2)}\right) = I^{(1)} + I^{(2)} - 2I^{(1)} I^{(2)}.
\end{equation}

Lastly, we take into account that both rods are in fact moving simultaneously. As suggested by Ref.~\cite{goujon2008mesoscopic}, four tests must be done: 1) a moving rod $\mathbf{AB}\rightarrow \mathbf{CD}$ intersects a stationary rod $\mathbf{PQ}$, 2) a moving rod $\mathbf{AB}\rightarrow \mathbf{CD}$ intersects a stationary rod $\mathbf{ST}$, 3) a moving rod $\mathbf{PQ}\rightarrow \mathbf{ST}$ intersects a stationary rod $\mathbf{AB}$, 4) a moving rod $\mathbf{PQ}\rightarrow \mathbf{ST}$ intersects a stationary rod $\mathbf{CD}$. The overall certainty of an intersection having taken place is then calculated by
\begin{equation}
I = I_1 I_2 \bar{I_3}\bar{I_4} + I_1 \bar{I_2} I_3 \bar{I_4} + \bar{I_1} I_2 \bar{I_3} I_4 + \bar{I_1} \bar{I_2} I_3 I_4,
\end{equation}
where $\bar{I} = \texttt{not}(I) = 1-I$ is the logical \texttt{not} operator.

Since the topological analysis is time consuming, we have only simulated a small system with $C=8$ chains having $N=16$ degrees of freedom, as depicted in Figure~\ref{3D}. It would be interesting to quantify the amount of chain crossings as a function of chain continuity parameter $J/N$, but unfortunately that would be an unfair comparison. The topological testing outlined above is prone to error near segment termini, and as we increase $J/N$ there are many more segments describing the same topology, and therefore the test becomes less valid.

However, we can compare the crossing rate as a function of the random force cutoff parameter $M$, plotted in Figure~\ref{certainty}. All bond pairs were examined at each of the $10^5$ time steps, and their crossing certainty score was binned into a histogram. We report the number of events exceeding the crossing score at levels of 1, 10, and 50\%. This number drops very sharply with $M$. One may expect a Boltzmann-like exponential decay $e^{-M}$, but the available data suggests that a power $M^{-4}$ falloff is more appropriate. For larger $M$ it is not entirely clear whether the few detected crossings are actual topological violations, or whether they can be attributed to the imperfection of the analysis method itself. We have visually inspected the configuration using a rotatable 3D plot, and could not confirm the analytically reported crossings. On the other hand, it is not impossible that some true chain crossings may have occurred during the $10^5$ steps and went unreported by the topological analysis. 

Either way, if we extrapolate the crossing rate with the help of the dashed line in the plot, then for a simulation with $M=120$ cutoff the crossing rate is $10^{-9}$. Scaling up to a bigger box with $C=64$ chains, $N=2048$ modes, we arrive at 1 crossing per $10^8$ time steps, per chain. This very crude estimate shows that some occasional crossings may have occurred, and their effect would be a small bias showing less entanglement than there should be.

\section{Conclusion and outlook}
In this paper we have explored a pseudo-continuous model of a polymer in semi-dilute solution, consisting of long repulsive chains whose motion is resolved using large Brownian time steps. By studying structural and mechanical correlations we have verified that the chains are indeed entangled and that their dynamical properties agree fairly well with the expected scaling laws. The model presented in this work is adequate to describe semi-dilute solutions at long time and distance scales. We have only assumed a linear spring attraction on the backbone and a soft Gaussian repulsive potential between the chains.

In closing, we emphasize that the goal of our simulation is not to prevent \emph{all} chain crossings, but only to \emph{reduce} their rate sufficiently for entanglement dynamics to emerge. The merit of a computer simulation is judged by various facets, including most importantly a realistic description of physics, but also the execution speed, the simplicity of the code, the number of assumptions and input parameters required, as well as its elegance and beauty. While no code can be perfect in all of these regards, we have tried to strike a suitable balance and we hope that our work will find many practical applications. These could include the study of polymer solutions under shear, in confined geometries, and using different chain architectures (star, comb, ring, brush), as well as heterogeneous polymer blends.

\begin{acknowledgments}
The authors thank Marcus M\"uller, Ralf Everaers, Giovanna Fragneto, and Felix Roosen-Runge for their useful comments on the draft version of this work. We also thank Anton Devishvili for programming help, as well as Mark Johnson and Luca Marradi for their assistance with the computational resources. Lastly, we acknowledge the use of \texttt{tubeplot} script by Janus H. Wesenberg to produce Figure~\ref{3D} in MATLAB.
\end{acknowledgments}

%\begin{suppinfo}
%\begin{enumerate}
%\item Visualisation of a single chain reptation.
%\item Appendix: Comparison with experimental data.
%\end{enumerate}
%\end{suppinfo}

\bibliography{manuscript}

\end{document}